\documentclass[sn-mathphys,Numbered]{sn-jnl}

\usepackage{graphicx}%
\usepackage{multirow}%
\usepackage{amsmath,amssymb,amsfonts}%
\usepackage{amsthm}%
\usepackage{mathrsfs}%
\usepackage[title]{appendix}%
\usepackage{xcolor}%
\usepackage{textcomp}%
\usepackage{manyfoot}%
\usepackage{booktabs}%
\usepackage{algorithm}%
\usepackage{algorithmicx}%
\usepackage{algpseudocode}%
\usepackage{listings}%

\theoremstyle{thmstyleone}%
\theoremstyle{thmstyletwo}%

\theoremstyle{thmstylethree}%

\begin{document}

\title[article title]{Spatial localization and diffusion of Dirac particles and waves induced by random temporal medium variations}


\author[1]{\fnm{Seulong} \sur{Kim}}

\author*[2,3]{\fnm{Kihong} \sur{Kim}}\email{khkim@ajou.ac.kr}

\affil[1]{\orgdiv{Research Institute of Basic Sciences}, \orgname{Ajou University}, \city{Suwon}, \postcode{16499}, \country{Korea}}

\affil[2]{\orgdiv{Department of Physics}, \orgname{Ajou University}, \city{Suwon}, \postcode{16499}, \country{Korea}}

\affil[3]{\orgdiv{School of Physics}, \orgname{Korea Institute for Advanced Study}, \city{Seoul}, \postcode{02455}, \country{Korea}}


\abstract
{
We investigate the consequences of temporal reflection on wave propagation and transformation in systems governed by a pseudospin-1/2 Dirac equation. These systems are spatially uniform but are subject to random temporal variations in mass, which correspond to the energy gap between the Dirac cones. By employing the invariant imbedding method on two complementary random models, we accurately compute all moments of temporal reflectance and derive their analytical expressions in short- and long-time regimes. In the long-time regime, the reflectance probability density is a constant equal to one, indicating uniform probability for any reflectance value. The group velocity of the wave decays to zero with time, signifying spatial localization induced by temporal variations. Numerical simulations of a wave pulse show that the initially narrow pulse evolves into a precisely Gaussian shape over time. In the long-time regime, the pulse center exhibits spatial localization, while its width shows ordinary diffusive behavior, increasing without limit. This behavior is universal, persisting regardless of the initial pulse shape or the probability distribution of the random mass. Our findings suggest that insulating behavior can be induced in Dirac materials by random temporal variations of the medium parameters. We discuss the possibilities of verifying our predictions in various experimental systems.
}




\maketitle

\section*{Introduction}\label{sec1}

When waves encounter spatial variations in medium properties, as described by second-order differential equations in space, they may scatter at spatial boundaries, producing backward-propagating waves. Now, consider a wave propagating through a medium undergoing temporal variations. Even when the medium is spatially uniform, the wave can scatter backward if it follows second-order differential equations in time, such as those describing electromagnetic (EM), acoustic, and water waves \cite{morg,fels,men,xiao,plan,bac,bac2}. Given the causality of time, these scattered waves travel in the opposite spatial direction. Although this phenomenon, known as temporal reflection, has long been recognized, there is currently a significant resurgence of interest in the propagation of waves in time-varying media \cite{cal1,eng,gali,most,pac,wang,zhou,li,mouss,dong}. This renewed interest is primarily driven by the motivation to develop innovative methods for controlling wave-matter interactions and creating versatile active metamaterials.


Many wave phenomena observed in spatially modulated systems find counterparts in temporally modulated cases. For instance, media with periodic temporal variations can exhibit a momentum gap \cite{zur,lust,kout}. Waves within this gap display exponential energy growth over time, contrasting with the exponential attenuation shown by waves within the frequency gap of spatially periodic media.
The present paper focuses on the influence of random temporal variations on wave propagation, a subject recently explored for EM and water waves by several research groups \cite{shar,carm,apf,garn,jkim}. In these studies, there is a consistent observation that wave energy increases exponentially over time. Theoretical investigations of EM waves have revealed a vanishing group velocity, indicating spatial localization akin to Anderson localization in spatially random media. Furthermore, the probability distribution of wave energy has been demonstrated to follow a log-normal distribution after a sufficiently long time \cite{shar,carm}.

Despite extensive studies of classical waves in time-varying media, research on analogous problems for quantum wave equations has been rare \cite{reck1}. One of the main reasons for this rarity is that the Schr\"odinger equation is of first order in time and does not exhibit any temporal reflection effect.
In our recent study, we investigated the propagation of waves governed by a generalized pseudospin-1/2 Dirac equation in the presence of various time-varying perturbations \cite{sk1}. These systems are effectively governed by a wave equation with a second-order derivative in time and exhibit a temporal reflection effect. Intriguingly, we demonstrated that momentum gaps do not appear in such systems under periodic temporal variations. Additionally, unlike EM waves, we established that exponential growth of total wave energy does not occur under any temporal variation.

Our primary objective in this work is to investigate the propagation and transformation of Dirac particles and waves under random temporal variations. Specifically, we explore cases in which the mass energy, governing the gap between the upper and lower Dirac cones, varies randomly over time.
We analyze two complementary models using the invariant imbedding method (IIM) \cite{sk2,sk3}. This method enables us to calculate all moments of the temporal reflectance, the probability density of the reflectance, and the spatial distribution of the wave fields in a numerically precise manner. Our findings reveal that the group velocity for plane waves and pulses decreases exponentially over time, indicating spatial localization. We derive analytical forms for all reflectance moments in short- and long-time regimes. Notably, we observe that in both regimes, the probability distributions follow the beta distribution, with the long-time regime demonstrating that the probability density is a constant equal to 1. This implies that the reflectance can take any value between 0 and 1 with equal probability.
Furthermore, we investigate the propagation of wave pulses with a nonzero central wave vector and finite pulse width through random temporal mass variations. Our findings reveal that the initially narrow pulse undergoes shape changes over time, transforming into a precisely Gaussian-shaped pulse. This pulse is a superposition of forward- and backward-propagating Gaussian pulses of identical shape. In the long-time regime, the pulse center exhibits spatial localization, while the pulse width displays diffusive behavior and increases without limit. This universal behavior persists irrespective of the initial pulse shape or the probability distribution of the random mass.

Our results can be directly tested through experiments on EM and acoustic waves in metamaterials designed specifically to exhibit Dirac-type dispersion \cite{mei,meta1,meta4,meta2}. In electronic systems that exhibit Dirac-type dispersion, such as graphene \cite{neto,weh,zhang,feng}, our findings indicate a fascinating phenomenon: {\it insulating} behavior, characterized by vanishing current and conductance, can be induced solely through the application of random temporal variations.

\section*{Models}\label{sec2}

We investigate the dynamics of waves governed by the pseudospin-1/2 Dirac equation in spatially uniform yet temporally varying media, with a focus on cases where only the effective mass of Dirac particles undergoes temporal variations. Due to spatial uniformity, the wave vector $\mathbf{k}$ remains a constant of motion. For simplicity, we assume that wave propagation occurs along the $x$ direction.

The time-dependent Dirac equation for the two-component vector wave function $\Psi=(\psi_1,\psi_2)^{\rm T}$ is expressed as
\begin{equation}
 i\hbar\frac{d}{dt}\Psi(t)=\begin{pmatrix}
  M(t)v_F^2 & \hbar k_x v_F  \\
  \hbar k_x v_F   & -M(t)v_F^2
 \end{pmatrix}\Psi(t),
 \label{eq:de}
 \end{equation}
 where $k_x$ represents the $x$ component of the wave vector, $M(t)$ is the time-dependent mass accounting for the energy gap between the upper and lower Dirac cones, and $v_F$ denotes the Fermi velocity. This equation can be transformed into
  \begin{equation}
 \ddot\psi_1+ikv_F \dot m\psi_1+\left(k v_F \right)^2\left(1+m^2\right)\psi_1=0,
 \end{equation}
where the dot denotes the derivative with respect to time, $k$ represents the wavenumber (magnitude of $k_x$), and $m$ is the normalized mass defined as $m=Mv_F/(\hbar k )$.
Since this is a second-order differential equation in time, it can exhibit temporal reflection under variations in the mass $m$ over time.
For a constant effective mass, the solution to this equation,
assuming a time dependence of $e^{-i\omega t}$, yields the dispersion relation
 $\omega=\pm  k v_F \sqrt{1+m^2}$.
The two solutions, with the plus and minus signs, correspond to particle-like and hole-like bands (denoted as $p$-band and $h$-band, respectively), representing the upper and lower Dirac cones.

\begin{figure}
  \centering
  \includegraphics[width=12cm]{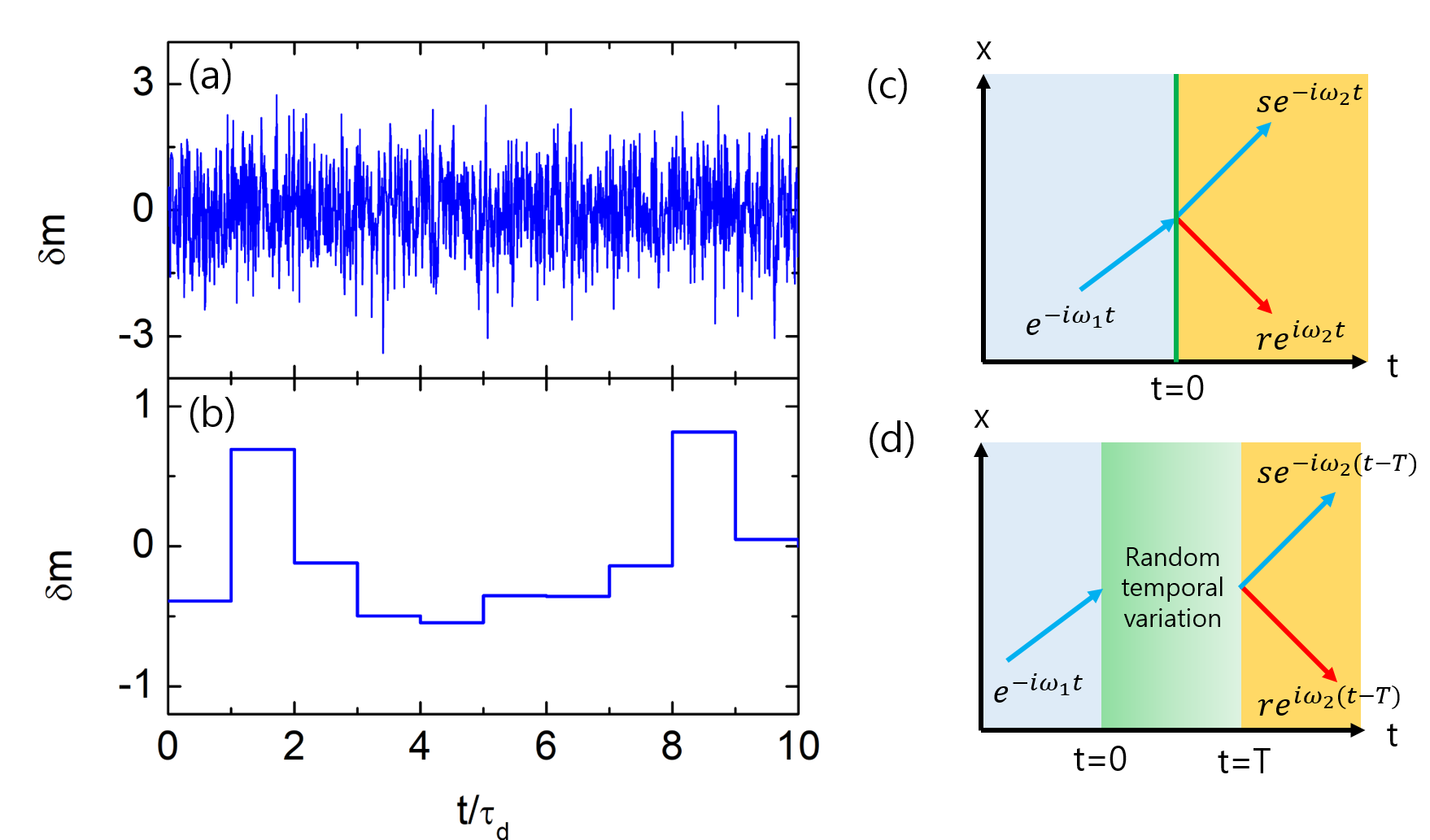}
  \caption{Specific realizations for (a) Model I, featuring $\delta$-correlated Gaussian disorder with $g_0 = 0.01$, and (b) Model II, characterized by stepwise uniformly distributed disorder with $\Delta m = 1$. (c) Scheme of temporal scattering at the temporal boundary where the medium properties vary, showing waves scattered into forward and backward waves. (d) Schematic of our model: from $t=0$ to $t=T$, the properties of the medium change randomly over time.}
  \label{fig1}
\end{figure}

 This study focuses on situations where the mass is a {\it random} function of time. We express the normalized mass $m(t)$ as the sum of an average value $m_0$ and a
 randomly varying function $\delta m(t)$:
 $m(t)=m_0+\delta m(t)$,
 where $\delta m(t)$ has a zero average and exhibits random behavior over time. We consider two supplementary random models for $\delta m(t)$,
 labeled as Model I and Model II. Model I adopts a $\delta$-correlated Gaussian random model, where $\delta m(t)$ satisfies
 \begin{equation}
 \langle\delta m(t)\delta m(t^\prime)\rangle=g_0\delta(t-t^\prime),
 ~ \langle\delta m(t)\rangle=0.
 \label{eq:delta}
 \end{equation}
 Here, $\langle\cdots\rangle$ denotes averaging over disorder, and $g_0$ characterizes the strength of disorder. In Model II,
 $\delta m$ remains constant during fixed time intervals ($\tau_d$), jumping to another value at subsequent intervals. These discontinuous variations persist for a total duration $T$, with each $\delta m$ value randomly selected from a uniform probability distribution within the range $[-{\Delta m},{\Delta m}]$. Model I is suitable for performing disorder-averaging analytically, while Model II can be used in more direct numerical simulations.
 In Figs.~\ref{fig1}(a) and \ref{fig1}(b), we illustrate specific realizations of disorder for Models I and II, respectively.
 This work employs the IIM to investigate the propagation of Dirac waves under these two random models and compare their behaviors.

\section*{Method}\label{sec3}

\subsection*{Invariant imbedding equations}

We consider a $p$-band wave with unit magnitude satisfying $\psi_1(t)=e^{-i\omega_1t}$ for $t\le 0$. The spatial dependence of all field components is represented by $e^{ikx}$ at all times.
Assuming temporal mass variations during the interval $0 \leq t \leq T$, the wave function $\psi_1$ before and after the interval is expressed as
\begin{equation}
 \psi_1(t,T)=\begin{cases}
         e^{-i\omega_1t}, & \mbox{if } t<0 \\
        r(T)e^{i\omega_2 (t-T)}+s(T)e^{-i\omega_2 (t-T)}, & \mbox{if } t>T
       \end{cases},
       \label{eq:drt}
 \end{equation}
where $\omega_1$ and $\omega_2$ are defined as $\omega_1 = kv_F\sqrt{1+m_1^2}$ and $\omega_2 = kv_F\sqrt{1+m_2^2}$, with $m_1$ and $m_2$ representing the normalized masses before and after the interval, respectively. The parameters $s$ and $r$ are the intraband and interband scattering coefficients, representing $\textit{p} \to \textit{p}$ and $\textit{p} \to \textit{h}$ transitions, respectively. These coefficients are considered functions of $T$.
In an isotropic medium, waves from different bands with the same momentum propagate in precisely opposite directions. Scattering caused by the time modulation of the mass does not alter the direction of the group velocity in each band \cite{sk1}. Therefore, $s$ and $r$ can be equivalently referred to as the (temporal) transmission and reflection coefficients, respectively. In Figs.~\ref{fig1}(c) and \ref{fig1}(d), we present schematics of temporal scattering and our model, respectively.

Using the IIM, we derive {\it exact} differential equations for $r$ and $s$ with respect to the imbedding parameter $\tau$, which denotes the total time interval over which the temporal variation occurs \cite{sm1}:
\begin{align}\
 &\frac{d}{d\tau}r(\tau)=ikv_F \left\{\frac{1+m(\tau)m_2}{\sqrt{1+m_2^2}} r(\tau)+[m(\tau)-m_2]\left(\frac{m_2}{\sqrt{1+m_2^2}}-1\right)s(\tau)\right\},\nonumber\\
  &\frac{d}{d\tau}s(\tau)=-ikv_F\left\{\frac{1+m(\tau)m_2}{\sqrt{1+m_2^2}} s(\tau)+[m(\tau)-m_2]\left(\frac{m_2}{\sqrt{1+m_2^2}}+1\right) r(\tau)\right\}.
 \label{eq:iie}
 \end{align}
 Interestingly, the same equations can be derived using an alternative method based on an intuitive heuristic approach \cite{sm22}.
 These equations are integrated from $\tau=0$ to $\tau=T$ using the initial conditions:
\begin{align}
r(0)=\frac{\sqrt{1+m_2^2}-m_2-\sqrt{1+m_1^2}+m_1}{2\sqrt{1+m_2^2}},~
s(0)=\frac{\sqrt{1+m_2^2}+m_2+\sqrt{1+m_1^2}-m_1}{2\sqrt{1+m_2^2}},
\label{eq:ic}
\end{align}
obtained either from the IIM \cite{sm1} or from the scattering coefficients at the temporal boundary between the two media with $m=m_1$ and $m=m_2$. When $m_1=m_2$, they reduce to $r(0)=0$ and $s(0)=1$. The reflectance and transmittance, defined as the ratios of the reflected and transmitted probability densities to the incident probability density, respectively, are expressed as:
\begin{align}
R=C_R \left\vert r\right\vert^2=\frac{1+m_2^2+m_2\sqrt{1+m_2^2}}{1+m_1^2-m_1\sqrt{1+m_1^2}}\left\vert r\right\vert^2,~
S=C_S \left\vert s\right\vert^2=\frac{1+m_2^2-m_2\sqrt{1+m_2^2}}{1+m_1^2-m_1\sqrt{1+m_1^2}}\left\vert s\right\vert^2.
\label{eq:rt}
\end{align}
The definitions of $C_R$ and $C_S$ are evident. Utilizing Eqs.~(\ref{eq:iie}) and (\ref{eq:rt}), we can demonstrate that the sum of $S$ and $R$, representing the total probability density of the particle, remains constant at 1 and is independent of $\tau$, regardless of the specific temporal variation of $m$ \cite{sm2}. This stands in sharp contrast to the case of EM waves in time-varying media, where the difference $(S-R)$ remains constant at 1, instead of the sum $(S+R)$.
Due to the constraint that $S+R=1$, both $S$ and $R$ should remain finite and not exhibit divergent behavior under any temporal variation.

\subsection*{Moments of the reflection and transmission coefficients}\label{sec33}

The IIM stands out as a powerful tool for solving the wave equation in random media, offering the unique capability to perform averaging over an infinite number of random configurations in an analytical manner. This is achieved by deriving nonrandom differential equations for ensemble-averaged moments.
In the case of Model I, which is a $\delta$-correlated random model
satisfying Eq.~(\ref{eq:delta}), we derive coupled nonrandom differential equations for moments denoted as
$Z_{abcd}\equiv\left\langle r^a{\left(r^{*}\right)^b}s^c\left(s^{*}\right)^d\right\rangle$ ($a,b,c,d=0,1,2,\cdots$) using Eq.~(\ref{eq:iie}) and Novikov's formula \cite{nov}.
For simplicity, we assume that the masses before and after the interval, as well as the average mass during the interval, are all equal (i.e., $m_1=m_2=m_0$). Then, the differential equations for the moments are expressed as \cite{sm3}
     \begin{eqnarray}
    &&\frac{1}{k v_F}\frac{d}{d\tau}Z_{abcd}=\bigg\{i\left(a-b-c+d\right)\sqrt{1+m_0^2}\nonumber\\&&~~~+\frac{g}{2}\left[\left(a+b+c+d+2ac+2bd\right)(B^2-1)-
(a-b-c+d)^2B^2\right]\bigg\} Z_{abcd}\nonumber\\
&&~~~+gc(a-b-c+d+1)B(B+1)Z_{a+1,b,c-1,d}\nonumber\\&&~~~-ga(a-b-c+d-1)B(B-1)Z_{a-1,b,c+1,d}\nonumber\\
    &&~~~-gd(a-b-c+d-1)B(B+1)Z_{a,b+1,c,d-1}\nonumber\\&&~~~+gb(a-b-c+d+1)B(B-1)Z_{a,b-1,c,d+1}\nonumber\\
    &&~~~+gcd(B+1)^2Z_{a+1,b+1,c-1,d-1}+gab(B-1)^2Z_{a-1,b-1,c+1,d+1}\nonumber\\
    &&~~~-gbc(B^2-1)Z_{a+1,b-1,c-1,d+1}
    -gad(B^2-1)Z_{a-1,b+1,c+1,d-1}\nonumber\\
    &&~~~-\frac{g}{2}c\left(c-1\right)\left(B+1\right)^2Z_{a+2,b,c-2,d}
    -\frac{g}{2}a\left(a-1\right)\left(B-1\right)^2Z_{a-2,b,c+2,d}\nonumber\\
    &&~~~-\frac{g}{2}d\left(d-1\right)\left(B+1\right)^2Z_{a,b+2,c,d-2}
    -\frac{g}{2}b\left(b-1\right)\left(B-1\right)^2Z_{a,b-2,c,d+2},
    \label{eq:mom}
  \end{eqnarray}
where the dimensionless disorder parameter $g$ and the constant $B$ are defined by
\begin{equation}
g=g_0kv_F,~B=\frac{m_0}{\sqrt{1+m_0^2}}.
\label{eq:defb}
\end{equation}
The initial conditions are $Z_{abcd}=1$ if $a=b=0$, and $Z_{abcd}=0$ for all other values.

The structure of Eq.~(\ref{eq:mom}) couples $Z_{abcd}$ to other moments $Z_{a^\prime b^\prime c^\prime d^\prime}$ that satisfy both $a^\prime+c^\prime=a+c$ and $b^\prime+d^\prime=b+d$. To obtain $Z_{nn00}$ [$=\langle R^n\rangle/(C_R)^n$] and $Z_{00nn}$ ($=\langle S^n\rangle$), only a finite number $[=(n + 1)^2]$ of coupled differential equations for $Z_{i,j,n-i,n-j}$ need to be solved, where $i$ and $j$ range from 0 to $n$ ($i,j=0,1,2,\cdots, n$). In this work, we investigate the propagation of Dirac waves by analyzing the behavior of $\left\langle R^n\right\rangle$.

\subsection*{Electric current density}

To connect with experimental observations, it is useful to derive an expression for the electric current density in terms of temporal transmittance and reflectance. The electric current density can be represented as $e J$, where $e$ is the charge and $J$ denotes the probability current. 
The continuity equation for the probability density $\rho$ of a Dirac particle, derived from Eq.~(\ref{eq:de}), is:
\begin{equation}
\frac{\partial}{\partial t}\rho+\frac{\partial}{\partial x}J=0,
\end{equation}
where
\begin{equation}
\rho=\Psi^{\dag}\Psi, \quad J=v_F\Psi^{\dag}\sigma_x\Psi =v_F\left(\psi_1^*\psi_2+\psi_2^*\psi_1\right).
\end{equation}
Using Eq.~(\ref{eq:drt}), we derive expressions for the current before and after a temporal variation as follows:
\begin{align}
&J_i= 2v_F\left(\sqrt{1+m_1^2}-m_1\right),  \\
&J(t > T) = 2v_F\left[\left(\sqrt{1+m_2^2}-m_2\right)\vert s\vert^2-\left(\sqrt{1+m_2^2}+m_2\right)\vert r\vert^2\right]\nonumber\\
&~~~~~~~~~~~~~~ - 4v_F m_2\left\{{\rm Re}(rs^*)\cos{[2\omega_2 (t-T)]} -{\rm Im}(rs^*) \sin{[2\omega_2 (t-T)]}\right\}.
\end{align}
The sinusoidal oscillating terms in $J(t>T)$ represent a standing wave and thus do not contribute to the net current.
From the definitions of $S$ and $R$ in Eq.~(\ref{eq:rt}), 
it becomes apparent that the ratio of the net current to the initial current is simply proportional to $S-R$:
\begin{equation}
\frac{J}{J_i} = \frac{\sqrt{1+m_1^2}}{\sqrt{1+m_2^2}}(S-R).
\label{eq:curr}
\end{equation}
This expression reduces to $S-R$ when $m_1 = m_2$, revealing a direct proportionality between the net current and the difference between temporal transmittance and reflectance. 

Thus far, we have focused on an effectively one-dimensional case where the Dirac cones are isotropic and the wave vector 
$\bf k$ possesses only an 
$x$ component, 
$k_x$. However, our method can be readily extended to higher-dimensional systems featuring anisotropic Dirac cones, where  
$\bf k$ includes components in all three spatial directions \cite{sm_as}. The results presented in the next section can be similarly applied to these cases with only minor adjustments. Furthermore, our method is applicable to semi-Dirac systems in two-dimensional lattices, where Dirac-like dispersion occurs in one direction and Schr\"odinger-like dispersion in another \cite{ban,huang,yee,oja}. In the most general semi-Dirac systems, the invariant imbedding equations differ from those presented here and require separate analysis, which represents an interesting direction for future work \cite{sm_as}.

\section*{Results}

\begin{figure}
  \centering
  \includegraphics[width=13cm]{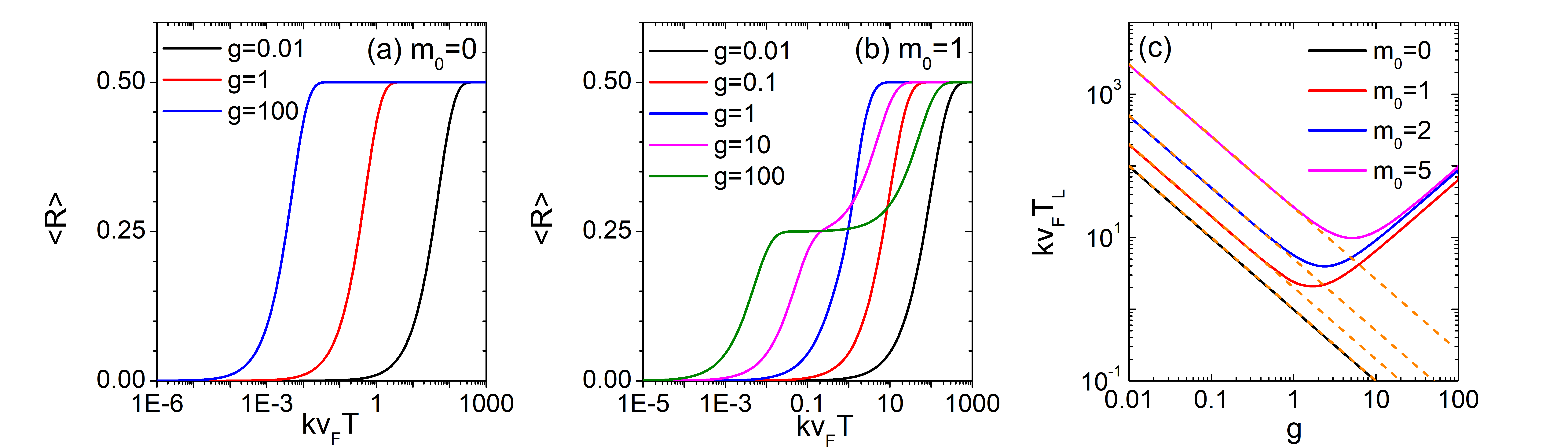}
  \caption{(a) and (b) Temporal reflectance, $\langle R \rangle$, averaged over temporal disorder in the Dirac mass for Model I, plotted as a function of the normalized time interval $kv_FT$. $\langle R \rangle$ is shown on a linear-log scale for various disorder strengths $g$, with average normalized mass parameters $m_0 = 0$ and $m_0 = 1$, respectively. (c) Normalized saturation time, $kv_FT_L$, defined as the time when $\langle R \rangle$ first exceeds $(1 - e^{-2})/2$,  plotted versus disorder strength $g$ on a log-log scale for $m_0 = 0$, $1$, $2$, and $5$. For $T > T_L$, both the average reflectance $\langle R \rangle$ and average transmittance $\langle S \rangle = 1 - \langle R \rangle$ stabilize at 0.5.}
  \label{f2}
\end{figure}

We now present the results from our numerical calculations using the IIM for both Model I and Model II. We begin our analysis with the simplest case in Model I, where the average mass $m_0$ and the masses $m_1$ and $m_2$ are set to zero in the presence of $\delta$-correlated Gaussian disorder. Under these conditions, the parameter $B$ in Eq.~(\ref{eq:mom}) becomes zero, resulting in the following expression:
\begin{eqnarray}
\frac{1}{k v_F}\frac{d}{d\tau}Z_{1100}=-2gZ_{1100}+g.
\label{eq:sic}
\end{eqnarray}
From this, we derive expressions for $\langle R\rangle$ ($=Z_{1100}$) and $\langle S\rangle$ ($=1-\langle R\rangle$):
\begin{align}
\langle R\rangle=\frac{1}{2}\left(1-e^{-2gkv_FT}\right),
\langle S\rangle=\frac{1}{2}\left(1+e^{-2gkv_FT}\right),
\label{eq:esrs}
\end{align}
where $T$ is the total interval of random temporal variation. The average reflectance $\langle R\rangle$ increases exponentially from zero, while the average transmittance $\langle S\rangle$ decreases exponentially from one. Both quantities saturate at 0.5 as $T$ exceeds $1/(gkv_F)$, as shown in Fig.~\ref{f2}(a). $\langle R\rangle$ represents the intensity for the \textit{h}-band state, which propagates with a group velocity of $-v_F$, whereas $\langle S\rangle$ represents the intensity for the \textit{p}-band state, which propagates with a group velocity of $v_F$. Consequently, the average energy $\langle E\rangle$, average propagation velocity $\langle v\rangle$, and average displacement $\langle x\rangle$ of the wave are expressed as functions of time:
\begin{align}
&\langle E\rangle=\hbar kv_F\sqrt{1+m_0^2}\left(\langle S\rangle-\langle R\rangle\right)=\hbar kv_Fe^{-2gkv_FT},\nonumber\\
&\langle v\rangle=v_F\left(\langle S\rangle-\langle R\rangle\right)=v_Fe^{-2gkv_FT},\nonumber\\
&\langle x\rangle=\int^T_0 \langle v(\tau)\rangle\,d\tau=\frac{1}{2gk}\left(1-e^{-2gkv_FT}\right).
\label{eq:zmass}
\end{align}
As $T$ becomes sufficiently large, the average energy and velocity of the wave approach zero. In other words, the total wave energy reaches an equilibrium value between the energies of the $p$- and $h$-bands and the wave ceases its movement. This is in sharp contrast to the case of EM waves, where the total wave energy increases exponentially over time \cite{shar,carm}. The average displacement approaches a constant value, given by $1/(2gk)$, which is inversely proportional to $g$, signifying spatial localization even in the absence of spatial randomness.
This type of spatial localization, induced solely by temporal random variations, is sometimes referred to as temporal Anderson localization.

In more general cases of Model I, where $m_0$ is nonzero, deriving exact analytical expressions valid at all times is challenging. Therefore, we numerically solve Eq.~(\ref{eq:mom}) to obtain all moments of the reflectance as a function of the time interval $T$. In Fig.~\ref{f2}(b), we present the average reflectance $\langle R\rangle$ when $m_0=1$ for different values of the disorder strength $g$. As $g$ increases, the behavior changes qualitatively. Initially, the average reflectance increases exponentially with time, then remains at some intermediate value
for a certain duration, and finally saturates at 0.5 above a sufficiently large value of $T$.
We define $T_L$ as the time interval beyond which $\langle R\rangle$ is greater than $(1-e^{-2})/2$ or exceeds $86.5\%$ of its final value. In Fig.~\ref{f2}(c), we depict $kv_FT_L$ versus $g$ on a log-log scale for various values of $m_0$. When $m_0$ is zero, $T_L$ decreases monotonically as $g$ increases, as indicated by Eq.~(\ref{eq:esrs}), where $kv_FT_L=1/g$. Conversely, when $m_0$ is nonzero, $T_L$ depends non-monotonically on $g$. For small $g$, $kv_FT_L\approx (1+m_0^2)/g$, while for large $g$, $kv_FT_L$ approaches $C g$, where $C$ is a constant of order 1 independent of $m_0$. $T_L$ is an experimentally relevant parameter that defines the long-time regime.

\begin{figure}
  \centering
  \includegraphics[width=9cm]{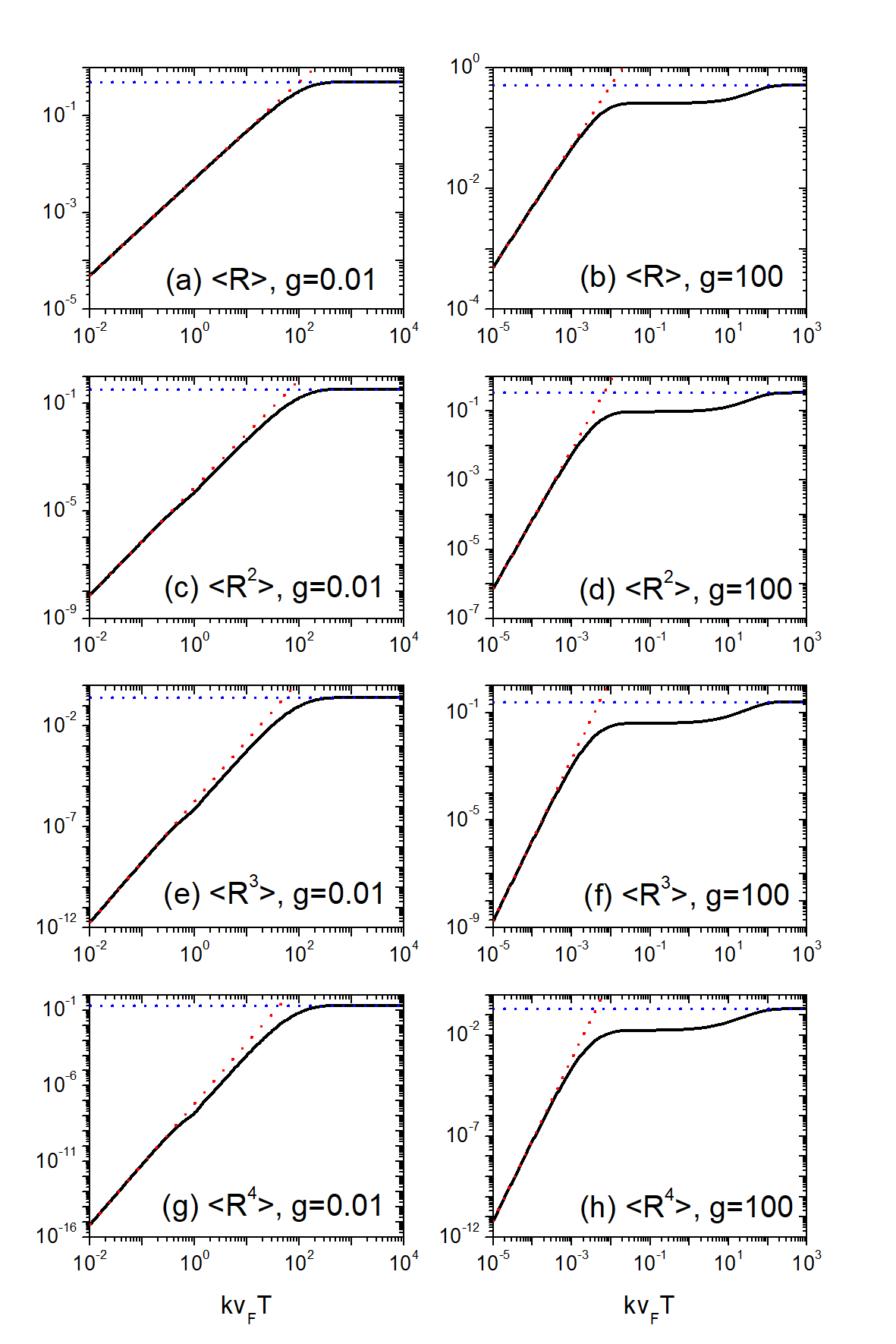}
  \caption{First four moments of temporal reflectance, $\left\langle R\right\rangle$, $\left\langle R^2\right\rangle$, $\left\langle R^3\right\rangle$, and $\left\langle R^4\right\rangle$, for Model I are plotted against
  the normalized time interval $kv_FT$ on a log-log scale. These are shown for $m_0=1$ with disorder parameters $g=0.01$ and 100. The red and blue dashed lines represent the analytical expressions from Eqs.~(\ref{eq:stlim}) and (\ref{eq:ltlim}) corresponding to the short- and long-time regimes, respectively. There is excellent agreement between the numerical results and the analytical predictions across all moments for both regimes.}\label{fig_mom}
\end{figure}

Remarkably, analytical forms for $\langle R^n\rangle$ can be derived in both the short-time regime, where $gkv_FT/(1+m_0^2)\ll 1$, and the long-time regime, where $T> T_L$ \cite{sm4}. In the former, we obtain
\begin{equation}
\langle R^n\rangle\approx(2n-1)!!\left(\frac{gkv_FT}{1+m_0^2}\right)^n,
\label{eq:stlim}
\end{equation}
while in the latter, we have
\begin{equation}
\langle R^n\rangle\approx\frac{1}{n+1}.
\label{eq:ltlim}
\end{equation}
In Fig.~\ref{fig_mom}, we present the first four reflectance moments versus $kv_FT$ on a log-log scale, when $m_0=1$ and $g=0.01$ and 100. The black solid lines depict numerical solutions obtained by solving Eq.~(\ref{eq:mom}), while the red and blue dashed lines correspond to the analytical expressions given by Eqs.~(\ref{eq:stlim}) and (\ref{eq:ltlim}), respectively. Excellent agreement is observed between the numerical results and analytical expressions in both the short- and long-time regimes.

\begin{figure}
  \centering
  \includegraphics[width=9cm]{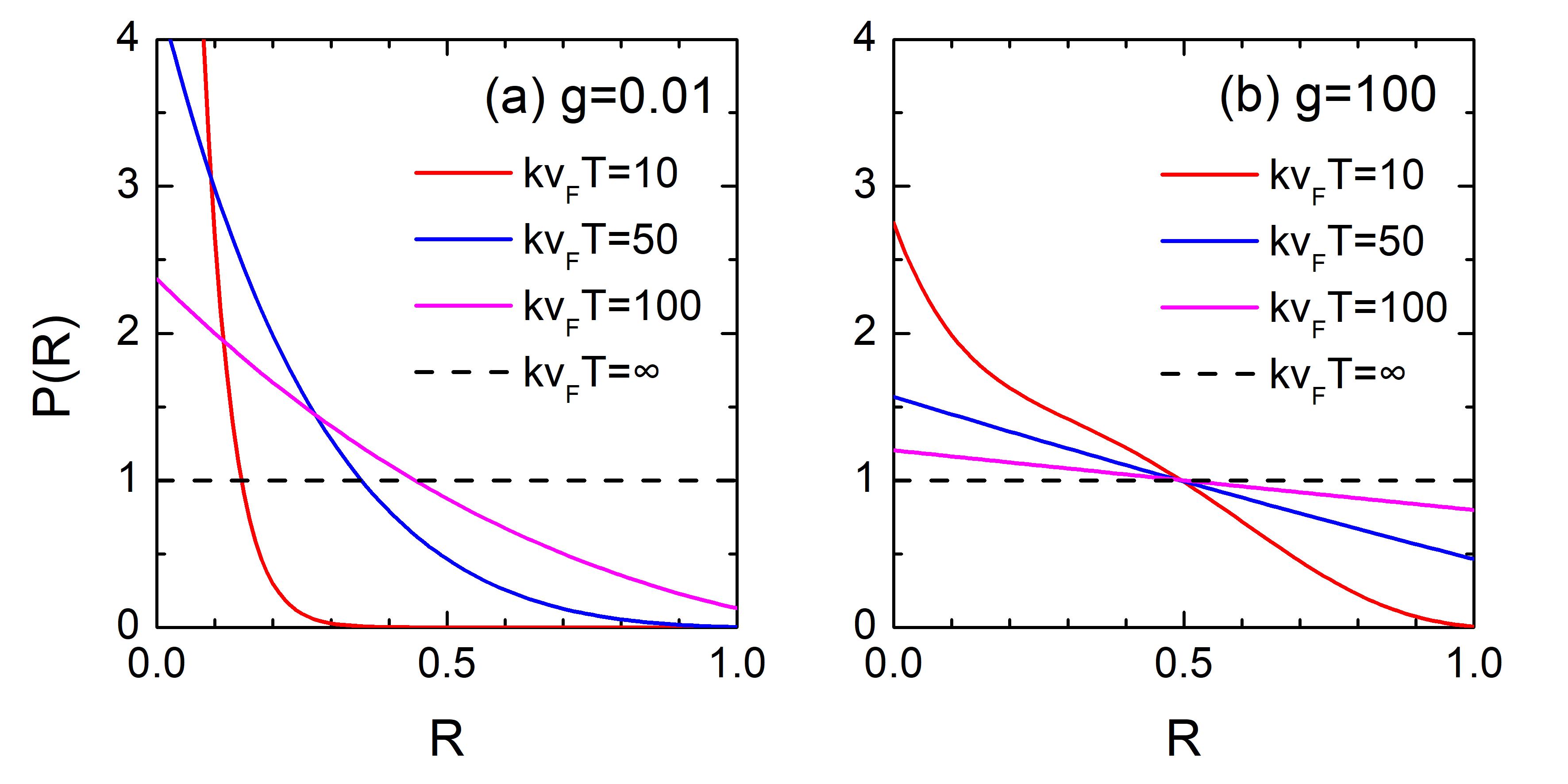}
  \caption{Evolution of the reflectance probability distribution function, $P(R)$, at normalized time intervals $kv_F T = 10$, $50$, and $100$ for Model I with $m_0 = 1$. Distributions are shown for disorder strengths $g = 0.01$ and $g = 100$. As time progresses, $P(R)$ converges to the uniform distribution $P(R) = 1$, with faster convergence observed at higher disorder strengths.}
  \label{fig_pdf}
\end{figure}

The probability distribution of the reflectance precisely follows the beta distribution in both the short- and long-time regimes \cite{sm5}.
In the long-time regime, $R$ follows a beta distribution of an extremely simple form with shape parameters $\alpha=\beta=1$, resulting in a uniform distribution, $P(R)=1$.
In other words, after a sufficiently long time has passed, the reflectance takes a random value from 0 to 1 with a uniform probability.
In the short-time regime, $\alpha=1/2$ and $\beta=(1+m_0^2)/(2gkv_FT)$ provide moments consistent with Eq.~(\ref{eq:stlim}). The probability density function $P(R)$ is calculated numerically by deducing all moments of the reflectance with high precision and utilizing the expansion of $P(R)$ in terms of shifted Legendre polynomials \cite{kim98}. In Fig.~\ref{fig_pdf}, we present $P(R)$ at different values of the normalized time interval for weak and strong disorder in Model I. As the time interval reaches and surpasses $T_L$, $P(R)$ gradually approaches the form of a uniform distribution, $P(R)=1$, with more rapid convergence observed for stronger disorder.

\begin{figure}
  \centering
   \includegraphics[width=9cm]{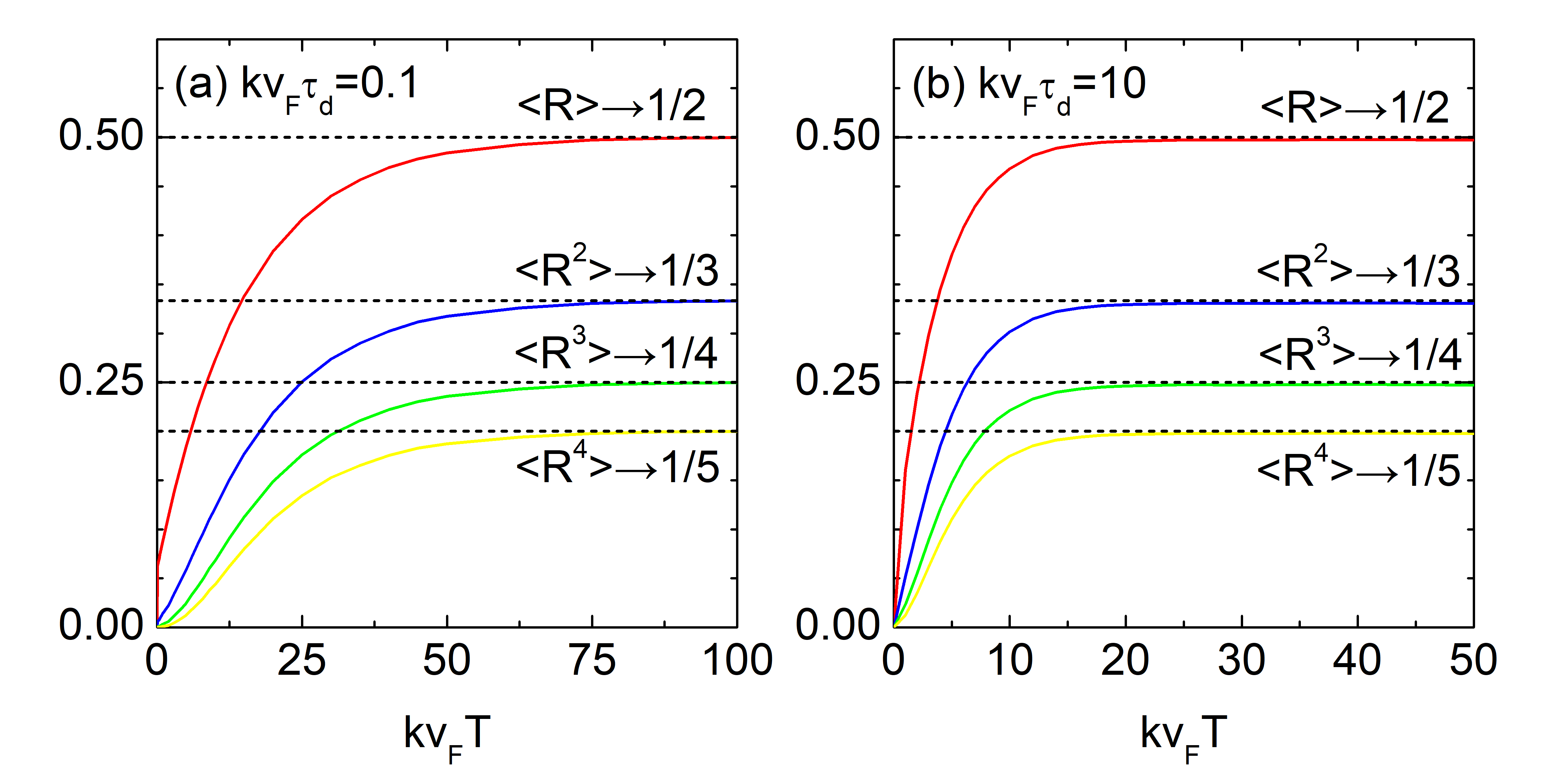}
   \includegraphics[width=9cm]{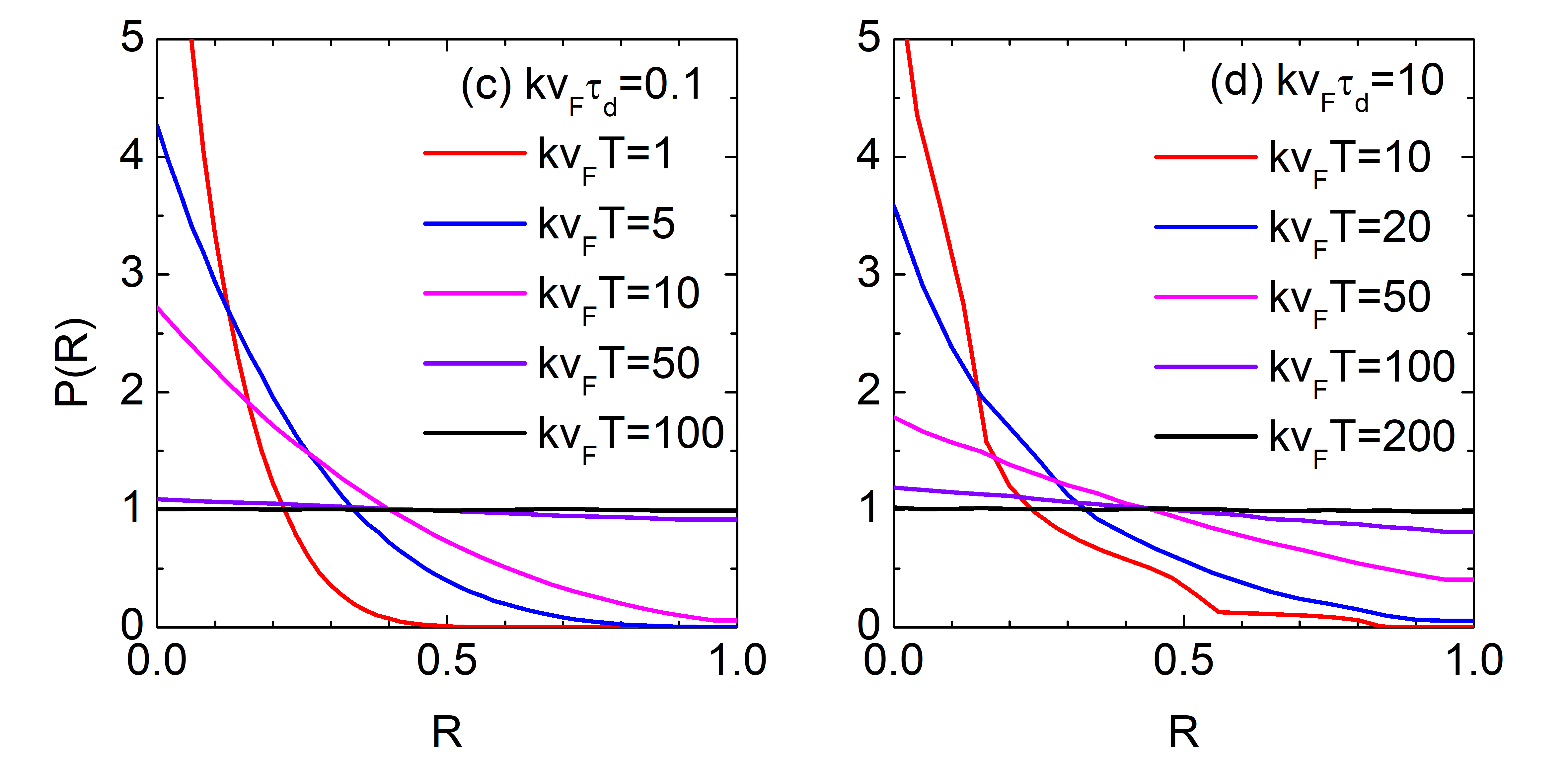}
  \caption{Panels (a) and (b) display the moments of reflectance in Model II as a function of $kv_F T$, with parameters $m_0 = 0$ and $\Delta m = 1$. In (a), $kv_F \tau_d = 0.1$, and in (b), $kv_F \tau_d = 10$. The moments are calculated by averaging over $10^6$ independent random configurations of $\delta m$. Dashed lines indicate the analytical expressions from Eq.~(\ref{eq:ltlim}) applicable in the long-time regime. Panels (c) and (d) illustrate the probability distribution of reflectance for different values of $T$ in Model II, with $m_0 = 0$, $\Delta m = 1$, and disorder parameters $kv_F \tau_d = 0.1$ in (c) and $kv_F \tau_d = 10$ in (d).}
  \label{f5}
\end{figure}

The investigation extends to Model II, characterized by the parameters $\Delta m$ and $\tau_d$. In this model, $\delta m$ is drawn from a uniform probability distribution within the range $[-\Delta m, \Delta m]$, remaining constant during a fixed time step $\tau_d$ and changing discontinuously during $N_s$ subsequent time steps until the total time interval $T$ ($=N_s\tau_d$) elapses. In Figs.~\ref{f5}(a) and \ref{f5}(b), we present reflectance moments in Model II, obtained by averaging over $10^6$ independent random configurations of $\delta m$, versus $kv_FT$, demonstrating that, as time progresses, the moments converge to $1/(n+1)$. Here, $m_0$ was set to zero for the sake of simplicity. Fitting the data for $\langle R\rangle$ to Eq.~(\ref{eq:esrs}) yields the effective disorder values of $g\approx 0.032$ for Fig.~\ref{f5}(a) and $g\approx 0.013$ for Fig.~\ref{f5}(b).
The corresponding effective values for $kv_FT_L$ are 31.3 and 76.9, respectively. In Figs.~\ref{f5}(c) and \ref{f5}(d), we illustrate the probability density $P(R)$ at different values of the normalized time interval in Model II, demonstrating convergence towards a uniform distribution as the time interval increases above $T_L$. This convergence is more rapid in cases of stronger effective disorder.

\begin{figure}
  \centering
  \includegraphics[width=13cm]{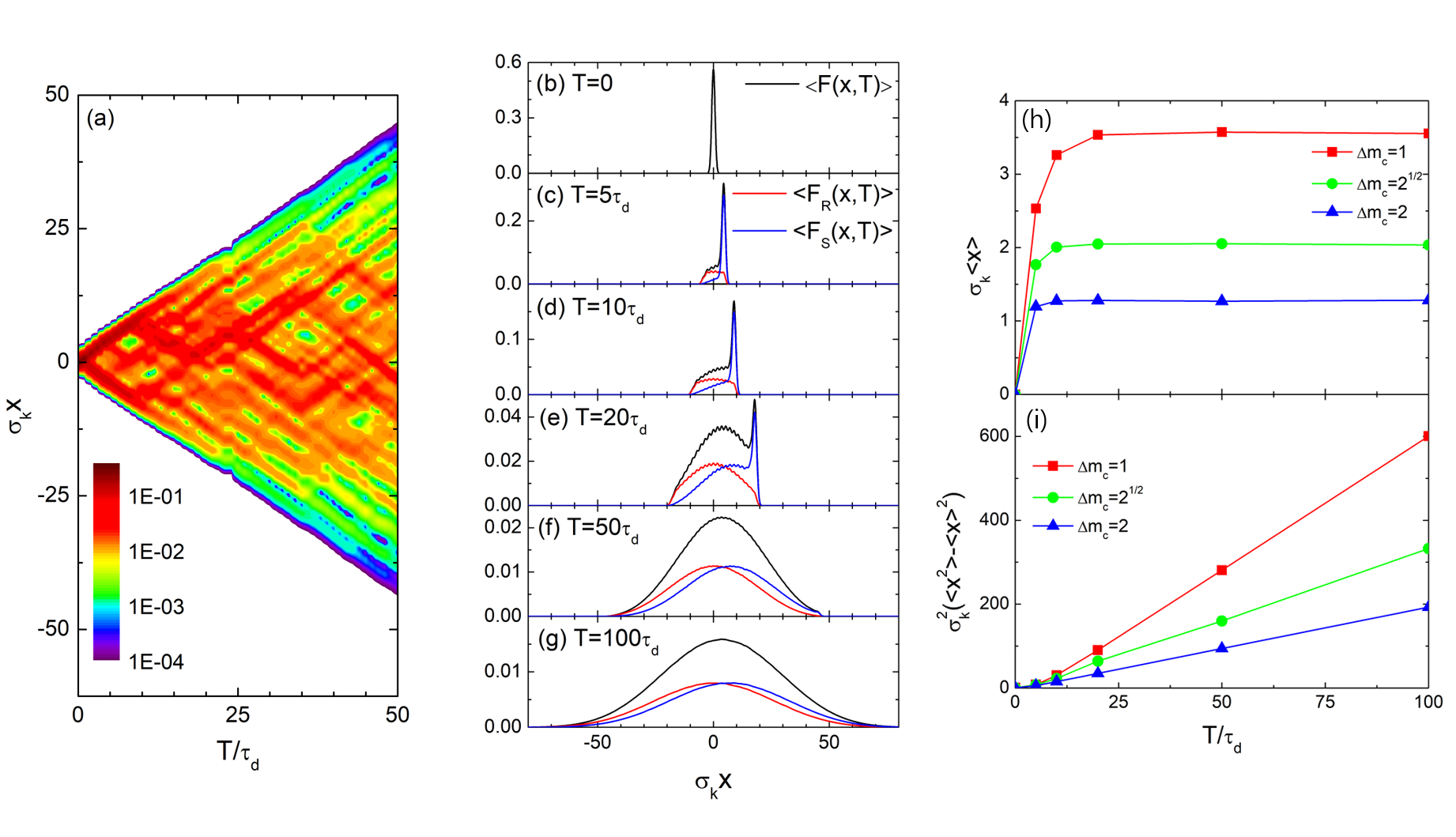}
  \caption{(a) Logarithmic contour plot of the normalized field intensity $F(x,T)$ showing the propagation and spread of a Gaussian pulse with central wave vector $k_c$ and pulse width $\sigma_k=0.1k_c$ in
  a single random configuration of mass in Model II. 
  Parameters are set as $m_0=0$, $\Delta m_c=1$, and $\Delta m(k_x)=k_c/\vert k_x\vert$.
  Only regions with $F(x,T)>10^{-4}$ are color-coded.
  The pulse, initially located at $x=0$, begins to propagate in the $+x$ direction.
  Time units are defined by $\tau_d=1/\left(\sigma_kv_F\right)$, and position $x$ is in units of $1/\sigma_k$. (b-g) Spatial distributions of $\left\langle F(x,T)\right\rangle$, $\left\langle F_R(x,T)\right\rangle$, and $\left\langle F_S(x,T)\right\rangle$ obtained by averaging over $10^5$ realizations of the propagating pulses, as considered in (a), at normalized time intervals $T/\tau_d=0$, 5, 10, 20, 50, and 100 for $m_0=0$ and $\Delta m_c=1$. (h) Mean displacement and (i) mean squared displacement of the pulse plotted against $T/\tau_d$ for $m_0=0$ with $\Delta m_c=1$, $\sqrt{2}$, and 2.}\label{fig_pulse}
\end{figure}

Our results thus far have provided a comprehensive understanding of wave behavior in both Model I and Model II, offering insights into the dynamic evolution of reflectance moments and probability distributions. These findings enhance our understanding of the complex interplay between temporal disorder and wave propagation. To facilitate a more direct comparison with experimental results, additional numerical simulations were conducted to explore how the evolving probability distribution affects the propagation of wave pulses with finite width.
Specifically, we examine the propagation of an initially Gaussian pulse characterized by a central wave vector $k_c$ ($>0$) and a pulse width $\sigma_k$ (set to $0.1k_c$), through a random temporal variation of mass over the total interval $T$. The initial pulse, comprising $p$-band states, is located at $x=0$ and begins to propagate in the $+x$ direction. For simplicity, the average mass $m_0$ is assumed to be zero, and the random temporal variation $\delta m$ is derived from Model II, slightly modified to reflect that the pulse consists of wave components with many
different values of $k_x$ in $-\infty<k_x<\infty$. The mass $M$ is randomly drawn from a uniform distribution within the range $[-\Delta M, \Delta M]$. At $k_x=k_c$, $\Delta m$ is set equal to $\Delta m_c$, defined as $\Delta m_c=v_F\Delta M/(\hbar k_c)$. For other $k_x$ values, $\Delta m(k_x)$ is scaled by $(k_c/|k_x|)\Delta m_c$. 

The Gaussian pulse is defined as
\begin{equation}
 u(x,t)=\int_{-\infty}^{\infty} D(k_x)e^{ik_xx}\Psi(k_x,t) dk_x,
 \label{eq:pul1}
\end{equation}
where $\Psi(k_x,t)$ for $t\ge T$ and $D(k_x)$ are given by
\begin{eqnarray}
&&\Psi(k_x,t)=
s(k_x,T)\begin{pmatrix} 1\\ \sqrt{1+m_2^2(k_x)}-m_2(k_x)\end{pmatrix} e^{-i\omega_2\left(t-T\right)}\nonumber\\
&&~~~~~~~~~~~+r(k_x,T)\begin{pmatrix} 1\\ -\sqrt{1+m_2^2(k_x)}-m_2(k_x)\end{pmatrix} e^{i\omega_2\left(t-T\right)},\nonumber\\
&&D(k_x)=e^{-\frac{\left(k_x-k_c\right)^2}{2\sigma_k^2}},
~\omega_2=\vert k_x\vert v_F\sqrt{1+m_2^2(k_x)},~m_2(k_x)=\frac{M_2v_F}{\hbar \vert k_x\vert}.
\label{eq:pul2}
\end{eqnarray}
Here, $M_2$ is chosen to be the same as the value of $M$ at the last time step for the given $T$. The transmission and reflection coefficients $s$ and $r$ are considered as functions of $k_x$ and $T$
in the present case.
The normalized field intensity $F(x,T)$ is obtained from
\begin{equation}
 F(x,T)=\frac{\left\vert u(x,T)\right\vert^2}{\int_{-\infty}^{\infty}\left\vert u(x,0)\right\vert^2 dx}.
 \label{eq:pul3}
\end{equation}

Under random temporal variations, the pulse undergoes repeated temporal reflections and broadens in space as it propagates. Eventually, the movement slows down, and the pulse center comes to a halt. The pulse transforms into a precisely Gaussian shape, and its width increases indefinitely, displaying ordinary diffusive behavior. In Fig.~\ref{fig_pulse}(a), we present a logarithmic contour plot of $F(x,T)$ for a single random configuration of $\Delta m(k_x)$ when $\Delta m_c=1$, where only the region with $F(x,T)>10^{-4}$ is depicted with colors. The time unit is defined by $\tau_d=1/(\sigma_kv_F)$, and the position $x$ is measured in units of $1/\sigma_k$. We observe that the pulse uniformly spreads throughout space.

In the current model, the effective disorder strength $g$ can be adjusted by varying $\Delta m_c$. We have determined that $g$ is approximately 0.013, 0.022, and 0.03 for $\Delta m_c$ values of 1, $\sqrt{2}$, and 2, respectively. In Figs.~\ref{fig_pulse}(b) to \ref{fig_pulse}(g), we show the spatial distribution of the pulse, averaged over $10^5$ realizations, at various values of $T$ for $\Delta m_c=1$. Numerically, we have found that the averaged intensity $\langle F(x,T)\rangle$ is composed of two components: $\langle F_S(x,T)\rangle$ and $\langle F_R(x,T)\rangle$, representing sub-pulses propagating in forward and backward directions, respectively.
We observe that when $T\le 20\tau_d$, the forward-propagating pulse consists of a coherent peak corresponding to  ballistic movement in the $+x$ direction and an incoherent part, which broadens and transforms into a Gaussian shape. Ultimately, the incoherent part dominates, and all of $\langle F(x,T)\rangle$, $\langle F_S(x,T)\rangle$, and $\langle F_R(x,T)\rangle$ take precisely Gaussian
shapes, while $\langle F_S(x,T)\rangle$ and $\langle F_R(x,T)\rangle$ become identical in shape.
Remarkably, we have verified that the averaged spatial distributions at long times are described universally by Gaussian functions, even when the mass probability distribution is not uniform or when the initial pulse shape is non-Gaussian \cite{sm6}.
We also note that when the random temporal variation is turned off after a sufficiently long duration, the forward- and backward-propagating
Gaussian pulses will separate. Then, by repeatedly turning the temporal variation on and off, it will be possible to generate many Gaussian pulses of identical shapes.

In Figs.~\ref{fig_pulse}(h) and \ref{fig_pulse}(i), we depict the mean displacement and mean squared displacement of the pulse obtained from $\left\langle F(x,T)\right\rangle$ as a function of the normalized interval $T/\tau_d$, when $m_0=0$ and $\Delta m_c=1$, $\sqrt{2}$, and 2.
According to Eq.~(\ref{eq:zmass}), the final position of the pulse should be close to $1/(2gk)$. Using the effective values of $g=0.013$, 0.022, and 0.03, we anticipate that the limiting values of $\sigma_k\langle x\rangle$ are 3.85, 2.27, and 1.67, respectively, which approximately agree with the numerically obtained values. Additionally, we observe that the mean squared displacement increases linearly as $T$ increases, displaying the behavior of ordinary diffusion. The diffusion coefficient defined by $D=(\langle x^2\rangle-\langle x\rangle^2)/(2T)$ decreases as the effective disorder increases. The average velocity of the pulse, defined by $d\langle x\rangle/dT$, decreases exponentially to zero, more rapidly in the presence of stronger disorder, displaying spatial localization.

\begin{figure}
  \centering
  \includegraphics[width=9cm]{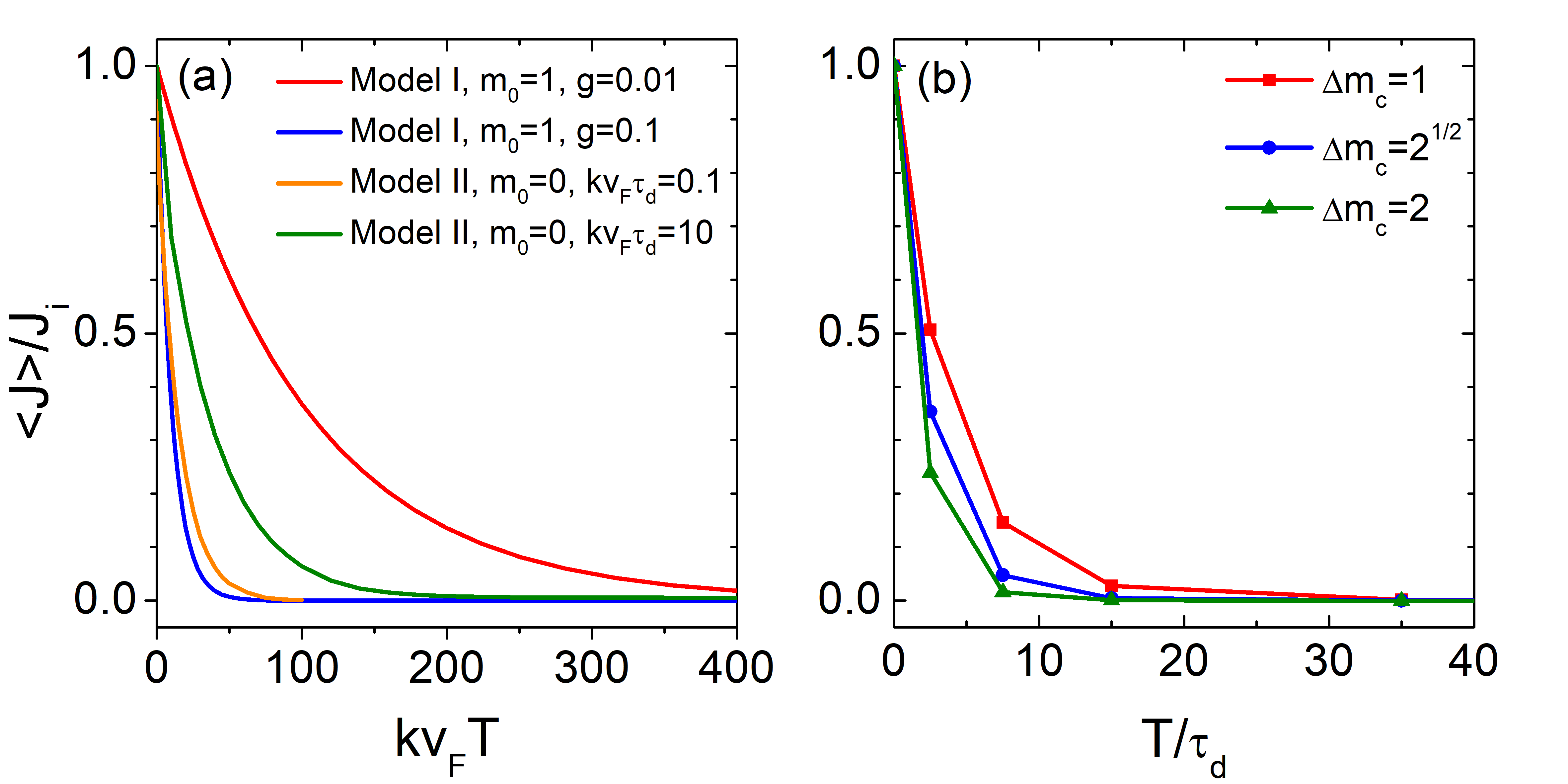}
  \caption{(a) The ratio of the averaged net current to the initial current, $\langle J\rangle/J_i$, is plotted against the normalized duration of the temporal variation, $kv_FT$. Results are shown for $m_0=1$
  with $g=0.01$ and $g=0.1$ in Model I, and for $m_0=0$ with $kv_F\tau_d=0.1$ and $kv_F\tau_d=10$ in Model II. (b) The ratio $\langle J\rangle/J_i$ for propagating Gaussian pulses,
  using parameters from Figs.~\ref{fig_pulse}(h) and \ref{fig_pulse}(i), with $m_0=0$ and $\delta m_c=1$, $\sqrt{2}$, and 2, is plotted against the normalized
  interval $T/\tau_d$.}
  \label{fig_cur}
\end{figure}

In Fig.~\ref{fig_cur}, we show the temporal variation of the ratio of the averaged net current to the initial current, as described by Eq.~(\ref{eq:curr}).
For $m_1=m_2$, this ratio simplifies to $\langle S\rangle-\langle R\rangle$. The figure presents results for different parameter values in Models I and II, as well as for propagating Gaussian pulses. In all cases,
$\langle J\rangle/J_i$ exhibits exponential decay, indicating
spatial localization induced by random temporal variations. The decay rate
increases with the effective disorder parameter. Although $\langle R\rangle$ for nonzero $m_0$, shown in Fig.~\ref{f2}(b), displays some nontrivial intermediate behavior, $\langle S\rangle-\langle R\rangle$ consistently follows a straightforward exponential decay pattern.

\section*{Discussion}

In this work, we investigated a novel phenomenon involving the spatial localization and diffusion of Dirac-type waves induced by random temporal variations in the energy gap between Dirac cones. Through rigorous analysis, we derived exact results and identified universal features governing the disorder-averaged propagation characteristics of both plane waves and pulses. These findings provide significant insights into wave-matter interactions, the development of active metamaterials, and the exploration of novel electronic properties in quantum materials.

A wide range of Dirac materials exhibit low-energy excitations that behave like Dirac particles \cite{neto,weh,zhang,feng}. It is possible to tune the medium parameters of these materials, such as mass, Fermi velocity, tilt velocity, as well as scalar and vector potentials \cite{sk1,kats,niz,jaur,gho,diaz,fara,mann,gap3,gap5,gap4,gap6,mendon}. For instance, in two-dimensional Dirac-type materials like silicene and germanene, the mass can be varied by adjusting the perpendicular electric field \cite{niz,jaur,gho} or by altering mechanical strain \cite{gap3,gap5,gap4,gap6}. Specifically, it has been estimated that an electric field of about 1 $\text {V/\AA}$ results in a band gap on the order of 0.1 eV \cite{niz}. While the induced gap size is not strictly proportional to the applied field strength, an approximate proportionality is expected to hold. Therefore, applying a randomly varying electric field over time is anticipated to produce a band gap magnitude corresponding to this proportionality.

We note that a temporally varying but spatially uniform modulation can induce backscattering in topological edge modes. Although topological edge states are typically robust against backscattering from static spatial disorder, time-dependent perturbations can introduce mechanisms for backscattering by enabling inelastic scattering processes and effectively breaking time-reversal symmetry. The phenomena predicted in this work may also be observed in topological insulators and superconductors, where the Dirac mass gap can be created or tuned through external parameters such as strain, temperature, gate voltage, or other controllable variables.

We anticipate that the effects explored in this study will manifest as insulating electronic behavior in these systems. The temporally induced spatial localization of Dirac electrons can be experimentally investigated by measuring electrical current and conductance. Introducing temporal disorder leads to a rapid decrease in electron group velocity and drift current, resulting in the net current approaching zero. From an experimental perspective, it is feasible to detect current attenuation due to temporal modulation, which could validate the theoretical predictions. Specifically, the time variation of the current should exhibit a form proportional to 
$S-R$. Moreover, when a pulse-like current is introduced, it encounters impedance due to localization while also dispersing diffusively. This process can be directly observed through techniques such as scanning ultrafast electron microscopy \cite{suem1,suem2,suem3}.

This phenomenon is not limited to electronic Dirac materials but can also occur in systems like cold atoms in optical lattices \cite{garr}, as well as Dirac-type photonic and acoustic metamaterials where electromagnetic and acoustic waves satisfy Dirac-type wave equations \cite{mei,meta1,meta4,meta2}. Research into dynamically altering the properties of metamaterials by changing their structure over time is ongoing \cite{tune2,tune3}. For example, in Dirac acoustic metamaterials, it is possible to dynamically tune the gap using piezoelectric material circuits \cite{tune1}. Similarly, adjusting the parameters of cold atomic systems in optical lattices using optical methods should be feasible. In photonic and acoustic metamaterials, it is easier to directly measure the transmitted and reflected waves, making the observation of such phenomena more straightforward.



\section*{Acknowledgments}
This research was supported through a National Research
Foundation of Korea Grant (NRF-2022R1F1A1074463)
funded by the Korean Government.
It was also supported by the Basic Science Research Program through the National Research Foundation of Korea funded by the Ministry of Education (NRF-2021R1A6A1A10044950).

\newpage
\setcounter{equation}{0}
\renewcommand{\theequation}{S\arabic{equation}}
\setcounter{figure}{0}
\renewcommand{\thefigure}{S\arabic{figure}}

\title{\Large Supplementary Material to ``Spatial localization and diffusion of Dirac particles and waves induced by random temporal medium variations''}
\maketitle

\section{Mathematical derivation of the invariant imbedding equations, Eq.~(5)}
\label{app:iie}

The invariant imbedding method (IIM) is a mathematical technique for solving boundary value problems involving multiple coupled ordinary differential equations (ODEs) 
by transforming them into equivalent initial value problems \cite{ssk2,ssk3}. In the context of wave equations in spatially inhomogeneous media, 
the IIM facilitates the calculation of reflection and transmission coefficients for a stratified medium with arbitrary inhomogeneity and thickness 
$L$ when a wave is incident from an external homogeneous region.
In its simplest form, the IIM transforms the original set of equations into an equivalent set of ODEs where the independent variable 
$l$ represents the thickness of the inhomogeneous slab rather than the position within it. For example, 
$r(l)$ denotes the reflection coefficient for a medium of thickness 
$l$, with the region 
$l<z\le L$ truncated from the original medium of thickness 
$L$. The invariant imbedding equation for 
$r(l)$ connects it to 
$r(L)$ through an initial value problem.
This method can be readily adapted to the current problem, where an arbitrary temporal variation persists over a finite time interval 
$T$. In this case, the invariant imbedding equation for the temporal reflection coefficient 
$r(\tau)$ for 
$0<\tau< T$ relates it to the target value 
$r(T)$ through an initial value problem. In this section, we derive the invariant imbedding equations, Eq.~(5) in the main text, in a mathematically rigorous manner.
In the next section, we present an alternative derivation based on an intuitive heuristic approach.

To apply the IIM to Eq.~(4), it is advantageous to reexpress it as
\begin{equation}
 u_1(t,T)\equiv\frac{\psi_1(t,T)}{s(T)}=\begin{cases}
         \rho_i(T)e^{-i\omega_1t}, & \mbox{if } t<0 \\
        \rho_r(T)e^{i\omega_2 (t-T)}+e^{-i\omega_2 (t-T)}, & \mbox{if } t>T
       \end{cases},
       \label{eq:ddrt}
 \end{equation}
where
\begin{equation}
\rho_i(T)=\frac{1}{s(T)},~\rho_r(T)=\frac{r(T)}{s(T)}. 
\label{eq:defcc}
\end{equation}
We also introduce $u_2(t,T)$ defined by $\psi_2(t,T)/s(T)$.
The functions $u_1$ and $u_2$ satisfy the Dirac equation, Eq.~(1), which can be recast as
\begin{eqnarray}
    \frac{\partial}{\partial t}{\bf u}(t,T)=A(t){\bf u}(t,T),
    \label{eq:we11}
\end{eqnarray}
where
\begin{eqnarray}
{\bf u}(t,T)=\begin{pmatrix} u_1(t,T) \\ u_2(t,T)\end{pmatrix},~A(t)=-ikv_F\begin{pmatrix} m(t) & 1 \\ 1 & -m(t)\end{pmatrix}.
\label{eq:we110}
\end{eqnarray}
The boundary conditions at $t=0$ and $t=T$ are given by
\begin{align}
    &u_1(0,T)=\rho_i(T),~u_2(0,T)=\left(\frac{\omega_1}{kv_F}-m_1\right)\rho_i(T),\nonumber\\
    &u_1(T,T)=\rho_r(T)+1,~u_2(T,T)=-\left(\frac{\omega_2}{kv_F}+m_2\right)\left[\rho_r(T)+1\right]+\frac{2\omega_2}{kv_F},
\end{align}
which can be combined to yield
\begin{align}
    G{\bf S}+H{\bf R}={\bf v},
    \label{eq:we12}
\end{align}
where
\begin{align}
   & {\bf S}=\begin{pmatrix} u_1(0,T) \\ u_2(0,T)\end{pmatrix},~
    {\bf R}=\begin{pmatrix} u_1(T,T) \\ u_2(T,T)\end{pmatrix},\nonumber\\
   & G=\begin{pmatrix}
  \frac{\omega_1}{kv_F}-m_1 & -1 \\
  0 & 0
\end{pmatrix},~
H=\begin{pmatrix}
  0 & 0 \\
  \frac{\omega_2}{kv_F}+m_2 & 1
\end{pmatrix},~{\bf v}=\begin{pmatrix}
  0 \\
  \frac{2\omega_2}{kv_F}
\end{pmatrix}.
\label{eq:we111}
\end{align}
In the linear problems, it can be shown that {\bf R} and {\bf S} are also linear in $\bf v$ and satisfy
\begin{align}
    {\bf R}=\tilde R {\bf v},~{\bf S}=\tilde S {\bf v},
    \label{eq:we13}
\end{align}
where $\tilde R$ and $\tilde S$ are $2\times 2$ matrices \cite{ssk2,ssk3}. From these definitions, it follows that
\begin{align}
    \tilde R_{12}v_2=\rho_r+1,~\tilde S_{12}v_2=\rho_i,
    \label{eq:defrs}
\end{align}
where $v_2=2\omega_2/(kv_F)$.

For the boundary value problems described by Eqs.~(\ref{eq:we11}), (\ref{eq:we12}), and (\ref{eq:we13}), it can be shown, using the standard theory of the IIM \cite{ssk2,ssk3}, that the following equations are generally satisfied:
\begin{align}
\frac{d}{d\tau}{\tilde R}(\tau)=A(\tau){\tilde R}(\tau)-{\tilde R}(\tau)HA(\tau){\tilde R}(\tau),~
\frac{d}{d\tau}{\tilde S}(\tau)=-{\tilde S}(\tau)HA(\tau){\tilde R}(\tau).
\label{eq:we30}
\end{align}
These equations can be integrated from $\tau=0$ to $\tau=T$ using the initial conditions given by
\begin{align}
{\tilde R}(0)={\tilde S}(0)=\left(G+H\right)^{-1}.
\label{eq:we31}
\end{align} 
By substituting the explicit forms of $A(\tau)$, $H$, and $G$ into these equations and using Eq.~(\ref{eq:defrs}), we can straightforwardly derive the invariant imbedding equations for $\rho_i$ and $\rho_r$:
\begin{align}
 \frac{d}{d\tau}\rho_i(\tau)=&ikv_F\left\{\frac{1+m(\tau)m_2}{\sqrt{1+m_2^2}}+\left[m(\tau)-m_2\right]\left(\frac{m_2}{\sqrt{1+m_2^2}}+1\right)\rho_r(\tau)\right\}\rho_i(\tau),\nonumber\\
 \frac{d}{d\tau}\rho_r(\tau)=&ikv_F\Bigg\{2\frac{1+m(\tau)m_2}{\sqrt{1+m_2^2}}\rho_r(\tau)+\frac{m_2\left[m(\tau)-m_2\right]}{\sqrt{1+m_2^2}}\left[1+\rho_r^2(\tau)\right]\nonumber\\
 &-\left[m(\tau)-m_2\right]\left[1-\rho_r^2(\tau)\right]\Bigg\},
 \label{eq:rho8}
\end{align}
along with the initial conditions
\begin{align}
\rho_i(0)=&\frac{2\sqrt{1+m_2^2}}{\sqrt{1+m_2^2}+m_2+\sqrt{1+m_1^2}-m_1},\nonumber\\
\rho_r(0)=&\frac{\sqrt{1+m_2^2}-m_2-\sqrt{1+m_1^2}+m_1}{\sqrt{1+m_2^2}+m_2+\sqrt{1+m_1^2}-m_1}.
\end{align}
Using the relationships provided by Eq.~(\ref{eq:defcc}), we can derive
\begin{align}
 \frac{ds}{d\tau}=-s^2\frac{d\rho_i}{d\tau},~\frac{dr}{d\tau}=s\frac{d\rho_r}{d\tau}-rs\frac{d\rho_i}{d\tau},
 \label{eq:stw}
\end{align}
and the initial conditions for $r$ and $s$ as given by Eq.~(6).
By combining Eq.~(\ref{eq:stw}) with Eq.~(\ref{eq:rho8}), we finally obtain the invariant imbedding equations for $r$ and $s$ as given by Eq.~(5) in the main text.

\section{Heuristic derivation of the invariant imbedding equations, Eq.~(5)}

\label{app:iie22}

\begin{figure}
  \centering
  \includegraphics[width=12cm]{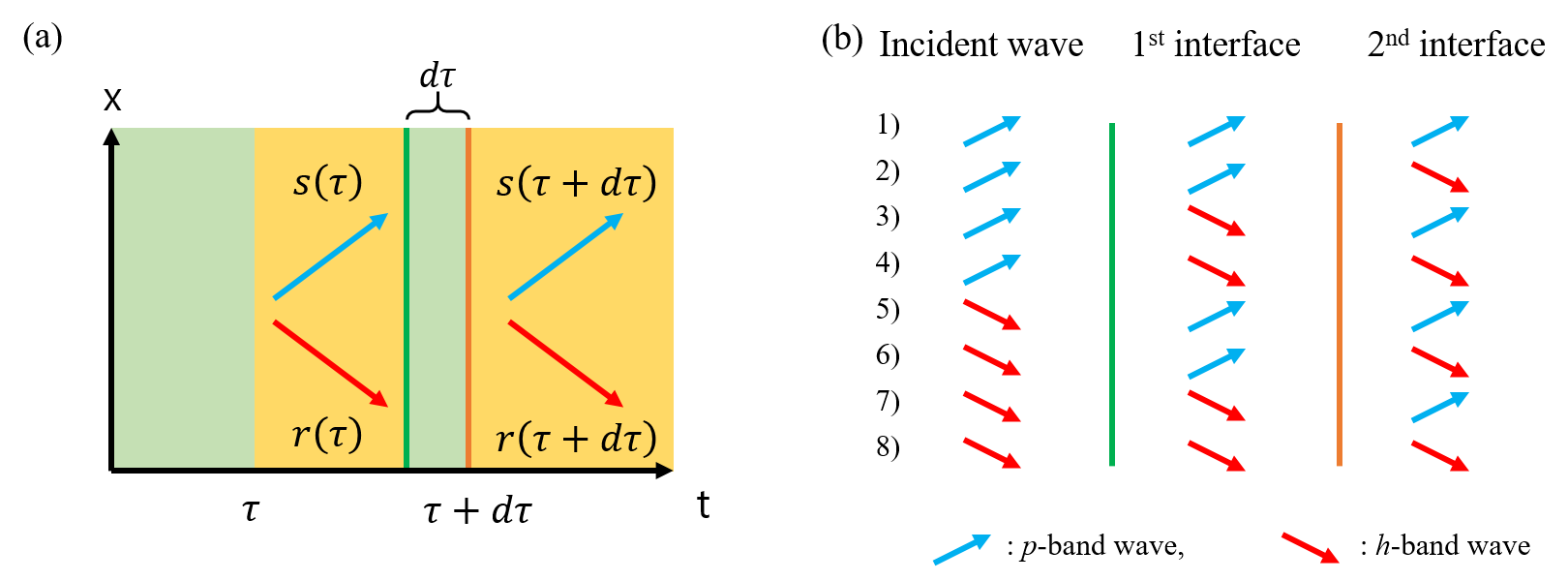}
  \caption{Schematic of temporal scattering processes. (a) The temporal transmission and reflection coefficients, $s(\tau+d\tau)$ and $r(\tau+d\tau)$, for variations over the interval $\tau+d\tau$, are expressed as linear combinations of $s(\tau)$ and $r(\tau)$. The coefficients of these combinations are determined by the temporal scattering coefficients within an infinitesimally narrow time slab $d\tau$. (b) Eight distinct scattering processes occurring within the temporal slab are illustrated.}
  \label{ggg2}
\end{figure}

In this section, we present an alternative derivation of the invariant imbedding equations, Eq.~(5), using an intuitive heuristic approach closely related to the transfer matrix method--a technique widely employed in optics and condensed matter physics. Indeed, the IIM can be considered a continuum version of the transfer matrix method.

The temporal scattering argument here is simpler than its spatial counterpart  because waves cannot propagate backward in time--they only move forward along the positive time direction ($+t$). This unidirectional propagation means that scattering at successive temporal boundaries can be described by straightforward multiplication of the new scattering coefficients with the previous ones. Consequently, the temporal transmission and reflection coefficients, $s(\tau+d\tau)$ and $r(\tau+d\tau)$, for temporal variations over an interval $\tau+d\tau$, can be obtained as linear combinations of $s(\tau)$ and $r(\tau)$, with coefficients expressed in terms of the temporal scattering coefficients over an infinitesimally narrow time slab $d\tau$.

The initial conditions given in Eq.~(6) of the main text, $s(0)$ and $r(0)$, correspond to the scattering coefficients when the mass parameter $m$ changes abruptly from $m_1$ to $m_2$ at time $t=0$.
We define $S_{pp}$ (equal to $s(0)$) as the intraband transition coefficient within the $p$-band and $S_{hp}$ (equal to $r(0)$) as the interband transition coefficient from the $p$-band to the $h$-band.
Similarly, $S_{hh}$ denotes the intraband transition coefficient within the $h$-band, and  
$S_{ph}$ represents the interband transition coefficient from the $h$-band to the $p$-band.

The explicit expressions for these scattering coefficients are \cite{ssk1}
\begin{align}
&S_{pp}=\frac{\sqrt{1+m_2^2}+m_2+\sqrt{1+m_1^2}-m_1}{2\sqrt{1+m_2^2}},~S_{hp}=\frac{\sqrt{1+m_2^2}-m_2-\sqrt{1+m_1^2}+m_1}{2\sqrt{1+m_2^2}},
\nonumber\\
&S_{hh}=\frac{\sqrt{1+m_2^2}-m_2+\sqrt{1+m_1^2}+m_1}{2\sqrt{1+m_2^2}},~
S_{ph}=\frac{\sqrt{1+m_2^2}+m_2-\sqrt{1+m_1^2}-m_1}{2\sqrt{1+m_2^2}}.
\end{align}
To derive 
$s(\tau+d\tau)$ and $r(\tau+d\tau)$ from $s(\tau)$ and $r(\tau)$, we consider
the addition of an infinitesimally narrow temporal slab between
$t=\tau$
and $t=\tau+d\tau$. The physical processes involved consist of three distinct steps:
\begin{enumerate}[leftmargin=1cm]
  \item First temporal boundary ($t=\tau$): The mass changes from $m_2$ to $m(\tau)$.
  \item Propagation through the time interval $d\tau$: Waves propagate freely during this infinitesimal interval.
  \item Second temporal boundary ($t=\tau+d\tau$): The mass changes from $m(\tau)$ back to $m_2$.
\end{enumerate}
We denote the scattering coefficients at the first interface with a single prime ($^\prime$) and those at the second interface with a double prime ($^{\prime\prime}$). The transmission and reflection coefficients after adding the temporal slab are then expressed as
\begin{align}
s(\tau+d\tau)=&s(\tau)\left(S_{pp}^\prime e^{-i\omega d\tau}S_{pp}^{\prime\prime} + S_{hp}^\prime e^{i\omega d\tau}S_{ph}^{\prime\prime}\right)\nonumber\\ 
&+r(\tau)\left(S_{ph}^\prime e^{-i\omega d\tau}S_{pp}^{\prime\prime} + S_{hh}^\prime e^{i\omega d\tau}S_{ph}^{\prime\prime}\right),\nonumber\\
r(\tau+d\tau)=&s(\tau)\left(S_{pp}^\prime e^{-i\omega d\tau}S_{hp}^{\prime\prime} + S_{hp}^\prime e^{i\omega d\tau}S_{hh}^{\prime\prime}\right)\nonumber\\
&+r(\tau)\left(S_{ph}^\prime e^{-i\omega d\tau}S_{hp}^{\prime\prime} + S_{hh}^\prime e^{i\omega d\tau}S_{hh}^{\prime\prime}\right),
\label{eq:zzz1}
\end{align}
where
$\omega = k v_F \sqrt{1 + m^2(\tau)}$.
In Fig.~\ref{ggg2}, we present a schematic diagram of the temporal scattering processes incorporated into these equations. Each of the eight distinct processes illustrated corresponds to a specific term in Eq.~(\ref{eq:zzz1}).

For the infinitesimal time interval $d\tau$, we can expand the exponential terms and the coefficients to first order:
\begin{align}
&e^{\pm i\omega d\tau}\approx 1 \pm i\omega d\tau,\nonumber\\
&s(\tau+d\tau)\approx s(\tau)+\frac{ds(\tau)}{d\tau}d\tau,\nonumber\\
&r(\tau+d\tau)\approx r(\tau)+\frac{dr(\tau)}{d\tau}d\tau.
\end{align}
Substituting these approximations into the expressions for $s(\tau+d\tau)$ and $r(\tau+d\tau)$ and retaining terms up to first order in $d\tau$, we obtain
\begin{align}
\frac{ds(\tau)}{d\tau}=&-i\omega \left(S_{pp}^\prime S_{pp}^{\prime\prime} - S_{hp}^\prime S_{ph}^{\prime\prime}\right)s(\tau) - i\omega \left(S_{ph}^\prime S_{pp}^{\prime\prime} - S_{hh}^\prime S_{ph}^{\prime\prime}\right)r(\tau),\nonumber\\
\frac{dr(\tau)}{d\tau}=&i\omega \left(S_{hp}^\prime S_{hh}^{\prime\prime}- S_{pp}^\prime S_{hp}^{\prime\prime}\right)s(\tau) + i\omega \left(S_{hh}^\prime S_{hh}^{\prime\prime} - S_{ph}^\prime S_{hp}^{\prime\prime}\right)r(\tau).
\label{eq:zzz2}
\end{align}
In deriving these equations, we have used the following identities:
\begin{align}
&S_{pp}^\prime S_{pp}^{\prime\prime} + S_{hp}^\prime S_{ph}^{\prime\prime}=S_{hh}^\prime S_{hh}^{\prime\prime} + S_{ph}^\prime S_{hp}^{\prime\prime}=1,\nonumber\\
&S_{ph}^\prime S_{pp}^{\prime\prime} + S_{hh}^\prime S_{ph}^{\prime\prime}=S_{hp}^\prime S_{hh}^{\prime\prime} + S_{pp}^\prime S_{hp}^{\prime\prime}=0.
\end{align}
Next, we compute the combinations of scattering coefficients appearing in the differential equations:
\begin{align}
&S_{pp}^\prime S_{pp}^{\prime\prime} - S_{hp}^\prime S_{ph}^{\prime\prime}=S_{hh}^\prime S_{hh}^{\prime\prime} - S_{ph}^\prime S_{hp}^{\prime\prime}= \frac{1 + m(\tau)m_2}{\sqrt{1 + m^2(\tau)} \sqrt{1 + m_2^2}},\nonumber\\
&S_{ph}^\prime S_{pp}^{\prime\prime} - S_{hh}^\prime S_{ph}^{\prime\prime}= \left[m(\tau) - m_2\right]\frac{m_2 + \sqrt{1 + m_2^2}}{\sqrt{1 + m^2(\tau)} \sqrt{1 + m_2^2}},\nonumber\\
&S_{hp}^\prime S_{hh}^{\prime\prime} - S_{pp}^\prime S_{hp}^{\prime\prime}= \left[m(\tau) - m_2\right]\frac{m_2 - \sqrt{1 + m_2^2}}{\sqrt{1 + m^2(\tau)} \sqrt{1 + m_2^2}}.
\end{align}
By substituting these expressions back into the differential equations, we arrive at Eq.~(5) of the main text, thus completing the alternative derivation. 

\section{Derivation of Eq.~(7) and the proof that $S+R=1$}
\label{app:rssum}

From Eq.~(4) and the relationship 
\begin{equation}
    \psi_2=\frac{i}{kv_F}\frac{\partial\psi_1}{\partial\tau}-m\psi_1,
\end{equation}
we obtain the expressions for $\psi_2$ before and after the interval:
\begin{equation}
 \psi_2(t,T)=\begin{cases}
         \left(\frac{\omega_1}{kv_F}-m_1\right) e^{-i\omega_1t}, & \mbox{if } t<0 \\
        -\left(\frac{\omega_2}{kv_F}+m_2\right) r(T)e^{i\omega_2 (t-T)}+\left(\frac{\omega_2}{kv_F}-m_2\right) s(T)e^{-i\omega_2 (t-T)}, & \mbox{if } t>T
       \end{cases}.
       \label{eq:sdrt}
 \end{equation}
 By taking the ratios of $\vert\psi_1\vert^2+\vert\psi_2\vert^2$ associated with the reflected and transmitted waves to that of the incident wave,
 we obtain 
 \begin{align}
R=\frac{1+\left(\frac{\omega_2}{kv_F}+m_2\right)^2}{1+\left(\frac{\omega_1}{kv_F}-m_1\right)^2}\vert r\vert^2,~
S=\frac{1+\left(\frac{\omega_2}{kv_F}-m_2\right)^2}{1+\left(\frac{\omega_1}{kv_F}-m_1\right)^2}\vert s\vert^2.
\end{align}
These equations can be readily reexpressed as Eq.~(7).

To demonstrate that $S+R=1$ at all times, we first calculate
  \begin{align}
  & \frac{d}{d\tau}\left(S+R\right)=C_S\left(s^*\frac{ds}{d\tau}+s\frac{ds^*}{d\tau}\right)+C_R\left(r^*\frac{dr}{d\tau}+r\frac{dr^*}{d\tau}\right)\nonumber\\
&~~~ =ikv_F\left(m-m_2\right)\left[C_S\left(\frac{m_2}{\sqrt{1+m_2^2}}+1\right)+C_R\left(\frac{m_2}{\sqrt{1+m_2^2}}-1\right)\right]\left(r^*s-rs^*\right)\nonumber\\
&~~~=0,
   \end{align}
where we have used Eqs.~(5) and (7). This shows that the sum of $S$ and $R$ remains constant over time. The sum is equal to 1
because
\begin{align}
   S(0)+R(0)= C_S\vert s(0)\vert^2+C_R\vert r(0)\vert^2=1.
\end{align}
This can be easily shown by substituting the expressions for $C_S$, $C_R$, $s(0)$, and $r(0)$ given by Eqs.~(6) and (7).

\section{Derivation of the differential equation for the moments, Eq.~(8)}
\label{app:z}

In the case where $m_1=m_2=m_0$ and $m=m_0+\delta m$, we can rewrite Eq.~(5) as
\begin{align}  
  &\frac{d}{d\tau}r(\tau)=ikv_F\left\{\sqrt{1+m_0^2}~r(\tau)+\left[Br(\tau)+\left(B-1\right)s(\tau)\right]\delta m(\tau)\right\},\nonumber\\  
  &\frac{d}{d\tau}s(\tau)=-ikv_F\left\{\sqrt{1+m_0^2}~s(\tau)+\left[Bs(\tau)+\left(B+1\right) r(\tau)\right]\delta m(\tau)\right\},
\end{align}
where $B$ is defined by Eq.~(9).
Then, the differential equation for the quantity $\tilde Z_{abcd}$ defined by $\tilde Z_{abcd}= r^a{\left(r^{*}\right)^b}s^c\left(s^{*}\right)^d$ becomes
\begin{align}
  &\frac{d}{d\tau}\tilde Z_{abcd}=ar^{a-1}{\left(r^{*}\right)^b}s^c\left(s^{*}\right)^d\frac{dr}{d\tau}
  +br^a{\left(r^{*}\right)^{b-1}}s^c\left(s^{*}\right)^d\frac{dr^*}{d\tau}\nonumber\\&~~~~~~~~~~~~~
  +cr^a{\left(r^{*}\right)^b}s^{c-1}\left(s^{*}\right)^d\frac{ds}{d\tau}
  +dr^a{\left(r^{*}\right)^b}s^c\left(s^{*}\right)^{d-1}\frac{ds^*}{d\tau}\nonumber\\
 &~~~~~ =ikv_F \left(a-b-c+d\right)\sqrt{1+m_0^2}~\tilde Z_{abcd}+ikv_F\left(a-b-c+d\right)B\tilde Z_{abcd}\delta m\nonumber\\
 &~~~~~~~ +ikv_F a\left(B-1\right)\tilde Z_{a-1,b,c+1,d}\delta m+ikv_F b\left(B-1\right)\tilde Z_{a,b-1,c,d+1}\delta m\nonumber\\
 &~~~~~~~ +ikv_F c\left(B+1\right)\tilde Z_{a+1,b,c-1,d}\delta m+ikv_F d\left(B+1\right)\tilde Z_{a,b+1,c,d-1}\delta m.
 \label{eq:zz0}
  \end{align}
Upon taking the average of both sides of this equation, we obtain  
\begin{align}
  &\frac{d}{d\tau} Z_{abcd}=ikv_F \left(a-b-c+d\right)\sqrt{1+m_0^2}~ Z_{abcd}+ikv_F\left(a-b-c+d\right)B\langle\tilde Z_{abcd}\delta m\rangle\nonumber\\
 &~~~~~~~~ +ikv_F a\left(B-1\right)\langle\tilde Z_{a-1,b,c+1,d}\delta m\rangle+ikv_F b\left(B-1\right)\langle\tilde Z_{a,b-1,c,d+1}\delta m\rangle\nonumber\\
 &~~~~~~~~ +ikv_F c\left(B+1\right)\langle\tilde Z_{a+1,b,c-1,d}\delta m\rangle+ikv_F d\left(B+1\right)\langle\tilde Z_{a,b+1,c,d-1}\delta m\rangle,
 \label{eq:zz1}
  \end{align}
  where $Z_{abcd}\equiv\langle \tilde Z_{abcd}\rangle$.
 Applying Nonikov's formula to Eq.~(\ref{eq:zz0}) \cite{snov}, we establish the following relationship:
  \begin{align}
  \langle \tilde Z_{abcd}\delta m\rangle
  =&\frac{ig}{2}\left(a-b-c+d\right)BZ_{abcd}+\frac{ig}{2}a\left(B-1\right)Z_{a-1,b,c+1,d}\nonumber\\
  &-\frac{ig}{2}b\left(B-1\right)Z_{a,b-1,c,d+1}-\frac{ig}{2}c\left(B+1\right)Z_{a+1,b,c-1,d}\nonumber\\
  &+\frac{ig}{2}d\left(B+1\right)Z_{a,b+1,c,d-1},
  \label{eq:zz2}
  \end{align}
 where $g$ is a dimensionless parameter defined by $g= g_0kv_F$. 
 By merging Eqs.~(\ref{eq:zz1}) and (\ref{eq:zz2}), we deduce the invariant imbedding equation for $Z_{abcd}$, as expressed in Eq.~(8), in a straightforward manner.

\section{Extending the invariant imbedding method to two-dimensional
systems with
anisotropic Dirac cones and semi-Dirac dispersions}
\label{app:iieg}

We first consider a two-dimensional Dirac model with anisotropic Dirac cones. We assume that the $x$ and $y$ components of the Fermi velocity, $v_{Fx}$ and $v_{Fy}$,
are generally different, and the wave vector $\bf k$ has both $x$ and $y$ components, $k_x$ and $k_y$. 
Then the Dirac equation corresponding to Eq.~(1) can be written as
\begin{equation}
 i\hbar\frac{d}{dt}\Psi(t)=\begin{pmatrix}
  M(t)v_F^2 & \hbar (k_x v_{Fx}-ik_yv_{Fy})  \\
  \hbar (k_x v_{Fx}+ik_yv_{Fy})   & -M(t)v_F^2
 \end{pmatrix}\Psi(t),
 \label{eq:sde}
 \end{equation}
 where $v_F$ is defined as $v_F=\sqrt{v_{Fx}^2+v_{Fy}^2}$.
 The derivation provided in Sec.~\ref{app:iie} can be generalized straightforwardly by replacing the matrices $A$, $G$, and $H$ given in Eqs.~(\ref{eq:we110}) and (\ref{eq:we111})
 with
\begin{align}
&A(t)=-ikv_F\begin{pmatrix} m(t) & \alpha \\ \alpha^* & -m(t)\end{pmatrix},\nonumber\\
  & G=\begin{pmatrix}
  \frac{\omega_1}{kv_F}-m_1 & -\alpha \\
  0 & 0
\end{pmatrix},~
H=\begin{pmatrix}
  0 & 0 \\
  \frac{\omega_2}{kv_F}+m_2 & \alpha
\end{pmatrix},
\label{eq:agh}
\end{align}
where $k$, $\alpha$, $m(t)$, $m_1$, $m_2$, $\omega_1$, and $\omega_2$ are defined as
\begin{align}
    &k=\sqrt{k_x^2+k_y^2},~\alpha=\frac{k_x v_{Fx}-ik_yv_{Fy}}{kv_F},\\
    &m(t)=\frac{M(t)v_F}{\hbar k},~
    m_1=\frac{M_1v_F}{\hbar k},~m_2=\frac{M_2v_F}{\hbar k},\label{eq:mmm}\\
    &\omega_1=kv_F\sqrt{\vert\alpha\vert^2+m_1^2},~\omega_2=kv_F\sqrt{\vert\alpha\vert^2+m_2^2}.
\end{align}
Following the same procedure as in Sec.~\ref{app:iie} and substituting $A$, $G$, and $H$ into Eqs.~(\ref{eq:we30}) and (\ref{eq:we31}), we obtain
the equations for $\rho_i$ and $\rho_r$:
\begin{align}
 \frac{d}{d\tau}\rho_i(\tau)=&ikv_F\left\{\frac{\left\vert\alpha\right\vert^2
 +m(\tau)m_2}{\sqrt{\left\vert\alpha\right\vert^2+m_2^2}}+\left[m(\tau)-m_2\right]
 \left(\frac{m_2}{\sqrt{\left\vert\alpha\right\vert^2+m_2^2}}+1\right)\rho_r(\tau)\right\}\rho_i(\tau),\nonumber\\
 \frac{d}{d\tau}\rho_r(\tau)=&ikv_F\Bigg\{2\frac{\left\vert\alpha\right\vert^2
 +m(\tau)m_2}{\sqrt{\left\vert\alpha\right\vert^2+m_2^2}}\rho_r(\tau)+\frac{m_2\left[m(\tau)-m_2\right]}
{\sqrt{\left\vert\alpha\right\vert^2+m_2^2}}\left[1+\rho_r^2(\tau)\right]\nonumber\\
 &-\left[m(\tau)-m_2\right]\left[1-\rho_r^2(\tau)\right]\Bigg\},
 \label{eq:rho}
\end{align}
with the initial conditions
\begin{align}
\rho_i(0)=&\frac{2\sqrt{\left\vert\alpha\right\vert^2+m_2^2}}
{\sqrt{\left\vert\alpha\right\vert^2+m_2^2}+m_2+\sqrt{\left\vert\alpha\right\vert^2+m_1^2}-m_1},\nonumber\\
\rho_r(0)=&\frac{\sqrt{\left\vert\alpha\right\vert^2+m_2^2}-m_2-\sqrt{\left\vert\alpha\right\vert^2+m_1^2}+m_1}
{\sqrt{\left\vert\alpha\right\vert^2+m_2^2}+m_2+\sqrt{\left\vert\alpha\right\vert^2+m_1^2}-m_1}.
\end{align}
Finally, the invariant imbedding equations for $r$ and $s$ can be rewritten as 
\begin{align}
 &\frac{d}{d\tau}r(\tau)=ikv_F \left\{\frac{\left\vert\alpha\right\vert^2+m(\tau)m_2}{\sqrt{\left\vert\alpha\right\vert^2+m_2^2}} r(\tau)+[m(\tau)-m_2]\left(\frac{m_2}{\sqrt{\left\vert\alpha\right\vert^2+m_2^2}}-1\right)s(\tau)\right\},\nonumber\\
  &\frac{d}{d\tau}s(\tau)=-ikv_F\left\{\frac{\left\vert\alpha\right\vert^2+m(\tau)m_2}{\sqrt{\left\vert\alpha\right\vert^2+m_2^2}} s(\tau)+[m(\tau)-m_2]\left(\frac{m_2}{\sqrt{\left\vert\alpha\right\vert^2+m_2^2}}+1\right) r(\tau)\right\}.
\end{align}
The corresponding initial conditions are
\begin{align}
&r(0)=\frac{\sqrt{\left\vert\alpha\right\vert^2+m_2^2}-m_2-\sqrt{\left\vert\alpha\right\vert^2+m_1^2}+m_1}
{2\sqrt{\left\vert\alpha\right\vert^2+m_2^2}},\nonumber\\
&s(0)=\frac{\sqrt{\left\vert\alpha\right\vert^2+m_2^2}+m_2+\sqrt{\left\vert\alpha\right\vert^2+m_1^2}-m_1}
{2\sqrt{\left\vert\alpha\right\vert^2+m_2^2}}.
\end{align}
By introducing the normalized mass $\tilde{m}=m/\left\vert\alpha\right\vert$, we obtain
\begin{align}
  &\frac{d}{d\tau}r(\tau)=i\left\vert\alpha\right\vert kv_F \left\{\frac{1+\tilde{m}(\tau)\tilde{m}_2}{\sqrt{1+\tilde{m}_2^2}} r(\tau)+[\tilde{m}(\tau)-\tilde{m}_2]\left(\frac{\tilde{m}_2}{\sqrt{1+\tilde{m}_2^2}}-1\right)s(\tau)\right\},\nonumber\\
  &\frac{d}{d\tau}s(\tau)=-i\left\vert\alpha\right\vert kv_F\left\{\frac{1+\tilde{m}(\tau)\tilde{m}_2}{\sqrt{1+\tilde{m}_2^2}} s(\tau)+[\tilde{m}(\tau)-\tilde{m}_2]\left(\frac{\tilde{m}_2}{\sqrt{1+\tilde{m}_2^2}}+1\right) r(\tau)\right\},
  \label{eq:iie00}
\end{align}
which have the same form as Eq.~(5) except for the redefined normalized mass and the overall constant factor $\vert\alpha\vert$.
This also allows the equation for the moments to adopt the identical form of Eq.~(8) by making the following substitutions:
\begin{align}
 m_0 \leftrightarrow \tilde m_0,~ g \leftrightarrow \frac{g}{\vert\alpha\vert},~ \tau \leftrightarrow \vert\alpha\vert \tau.
\end{align}
In other words, even if the medium is two-dimensional and anisotropic, the phenomena described in the main text remain the same, differing only quantitatively.

Next, we explore semi-Dirac systems in two-dimensional lattices characterized by Dirac-like dispersion in one direction and Schr\"odinger-like dispersion in the other. These systems are governed by the Hamiltonian:
\begin{equation}
  \mathit{H}=\begin{pmatrix}
  M(t)v_F^2 & \hbar k_x v_F-i\left(\frac{\hbar^2}{2\mu}k_y^2+\Delta\right)  \\
  \hbar k_x v_F+i\left(\frac{\hbar^2}{2\mu}k_y^2+\Delta\right)   & -M(t)v_F^2
 \end{pmatrix},
\end{equation}
with two types of gap parameters $M$ and $\Delta$ \cite{soja}. Assuming $k_x$ is positive for simplicity,
if the parameter $\Delta$ remains constant over time, the derivation of the invariant imbedding equations is similar to that for anisotropic Dirac cones.
By replacing $k$ with $k_x$ in
Eqs.~(\ref{eq:agh}) and (\ref{eq:mmm}--\ref{eq:iie00}), and defining $\alpha$ as
\begin{equation}
 \alpha=1-i\left(\frac{\hbar k_y^2}{2\mu k_xv_F}+\frac{\Delta}{\hbar v_Fk_x}\right),
\end{equation}
our main results are applicable in the same way to these semi-Dirac systems.
 
However, if $\Delta$ depends on time, the parameter $\alpha$ becomes time-dependent as well, and the invariant imbedding equations take different forms compared to the previous cases.
In this situation, we define $G$ and $H$ as
\begin{align}
   & G=\begin{pmatrix}
  \frac{\omega_1}{k_xv_F}-m_1 & -\alpha_1 \\
  0 & 0
\end{pmatrix},~
H=\begin{pmatrix}
  0 & 0 \\
  \frac{\omega_2}{k_xv_F}+m_2 & \alpha_2
\end{pmatrix},
\end{align}
where
\begin{eqnarray}
\omega_1=v_Fk_x\sqrt{\left\vert\alpha_1\right\vert^2+m_1^2},
~\omega_2=v_Fk_x\sqrt{\left\vert\alpha_2\right\vert^2+m_2^2}.
\end{eqnarray}
The parameters $\alpha_1$ and $\alpha_2$ are the values of $\alpha$ before and after the interval of temporal variation, respectively.
Then, the invariant imbedding equations adopt a different form:
\begin{align}
  &\frac{d}{d\tau}r(\tau)=i\frac{k_xv_F}{\sqrt{\left\vert\alpha_2\right\vert^2+m_2^2}}\left\{\left[m(\tau)m_2
  +\frac{\alpha^*(\tau)\alpha_2+\alpha(\tau)\alpha_2^*}{2}\right]r(\tau)\right.\nonumber\\
  &~\left.+\left[\frac{\alpha^*(\tau)\alpha_2-\alpha(\tau)\alpha_2^*}{2}
  +\left(m-\frac{m_2\alpha(\tau)}{\alpha_2}\right)\left(m_2-\sqrt{\left\vert\alpha_2\right\vert^2+m_2^2}\right)\right]s(\tau)\right\},\nonumber\\
  &\frac{d}{d\tau}s(\tau)=-i\frac{k_xv_F}{\sqrt{\left\vert\alpha_2\right\vert^2+m_2^2}}\left\{\left[m(\tau)m_2
  +\frac{\alpha^*(\tau)\alpha_2+\alpha(\tau)\alpha_2^*}{2}\right]s(\tau)\right.\nonumber\\
  &~\left.+\left[\frac{\alpha^*(\tau)\alpha_2-\alpha(\tau)\alpha_2^*}{2}
  +\left(m-\frac{m_2\alpha(\tau)}{\alpha_2}\right)\left(m_2+\sqrt{\left\vert\alpha_2\right\vert^2+m_2^2}\right)\right]r(\tau)\right\}.
\end{align}
These equations are integrated from $\tau=0$ to $\tau=T$ using the initial conditions:
\begin{align}
&r(0)=\frac{\alpha_1\left(\sqrt{\left\vert\alpha_2\right\vert^2+m_2^2}-m_2\right)
-\alpha_2\left(\sqrt{\left\vert\alpha_1\right\vert^2+m_1^2}-m_1\right)}
{2\alpha_1\sqrt{\left\vert\alpha_2\right\vert^2+m_2^2}},\nonumber\\
&s(0)=\frac{\alpha_1\left(\sqrt{\left\vert\alpha_2\right\vert^2+m_2^2}+m_2\right)
+\alpha_2\left(\sqrt{\left\vert\alpha_1\right\vert^2+m_1^2}-m_1\right)}
{2\alpha_1\sqrt{\left\vert\alpha_2\right\vert^2+m_2^2}}.
\end{align}
As these equations differ from previous cases, they require independent analysis, offering an interesting direction for future research.

\section{Derivation of the analytical expression of the reflectance moments in the short-time regime, Eq.~(18)}
\label{app:asyms}

In this and the following sections, we present the derivation of analytical formulas for both the short- and long-time regimes, providing insights into the behavior of the system across various time scales.
In the short-time regime, we expand the ensemble-averaged moment $Z_{abcd}$, where $a$, $b$, $c$, and $d$ are nonnegative integers, as a power series with respect to the very short time $T$:
\begin{equation}
Z_{abcd}=Z_{abcd}^{(0)}+Z_{abcd}^{(1)}T+Z_{abcd}^{(2)}T^2+Z_{abcd}^{(3)}T^3+\cdots.
\label{eq:zps}
\end{equation}
Here, $Z_{abcd}^{(j)}$ represents the coefficient of the \textit{j}-th order term in the expansion series of $Z_{abcd}$. By applying the initial condition at $T = 0$, we can determine the coefficients $Z^{(0)}_{abcd}$ of the zeroth order term:
\begin{equation}
Z_{abcd}^{(0)}=\begin{cases}
                1, & \mbox{if } a=b=0 \\
                0, & \mbox{otherwise}
              \end{cases}.
\label{eq:zps1}
\end{equation}
The higher-order coefficients $Z_{abcd}^{(j)}$ with $j\ge 1$ can be obtained by substituting Eq.~(\ref{eq:zps}) into Eq.~(8) and applying the perturbation method.
Due to the constraint imposed by the zeroth-order term, as specified in Eq.~(\ref{eq:zps1}), we observe that only the three terms proportional to $(B-1)^2$ in Eq.~(8) are pertinent for determining the moments $Z_{nn00}$.
In particular, we find that the relationship of the form
\begin{align}
Z_{abcd}^{(j)}=\frac{\gamma}{j}&\left[ab Z_{a-1,b-1,c+1,d+1}^{(j-1)}-\frac{a(a-1)}{2}Z_{a-2,b,c+2,d}^{(j-1)}
-\frac{b(b-1)}{2}Z_{a,b-2,c,d+2}^{(j-1)}\right],
\label{eq:recursion}
\end{align}
where
\begin{equation}
\gamma=gkv_F(B-1)^2,
\end{equation}
can be used in the determination of the coefficients.
We note that this equation does not have an imaginary term, and therefore, the solutions for $Z_{abcd}^{(j)}$ are always real. 

Using Eq.~(\ref{eq:recursion}) in conjunction with Eq.~(\ref{eq:zps1}), we can derive the first-order term of $Z_{1100}$ as follows:
\begin{align}
Z_{1100}^{(1)}&=\gamma Z_{0011}^{(0)}=\gamma,
\end{align}
and consequently, 
\begin{align}
\langle R\rangle&\approx \gamma C_R=\frac{gkv_FT}{1+m_0^2}.
\end{align}
By utilizing Eqs.~(\ref{eq:zps1}) and (\ref{eq:recursion}), it is straightforward to demonstrate that when calculating $Z_{nn00}$ with $n\ge 2$, all coefficients $Z_{nn00}^{(j)}$ with $j<n$ vanish.
In other words, the leading term of $Z_{nn00}$ is of order $n$.
We can express the second-order coefficient for $Z_{2200}$ as follows:
\begin{equation}
Z_{2200}^{(2)}=\frac{\gamma}{2}\left(4Z_{1111}^{(1)}-Z_{0220}^{(1)}
-Z_{2002}^{(1)}\right).
\end{equation}
Again using Eqs.~(\ref{eq:zps1}) and (\ref{eq:recursion}), we easily verify that $Z_{1111}^{(1)}=\gamma$ and $Z_{0220}^{(1)}=Z_{2002}^{(1)}=-\gamma$. 
Then, we obtain $Z_{2200}^{(2)}=3\gamma^2$ and
\begin{align}
\langle R^2\rangle&\approx 3\gamma^2 C_R^2=3!!\left(\frac{gkv_FT}{1+m_0^2}\right)^2,
\end{align}
where we have used $3!!=3$ in anticipation of the general result.

For the third-order coefficient of $Z_{3300}$, we have
\begin{equation}
Z_{3300}^{(3)}=\frac{\gamma}{3}\left(9Z_{2211}^{(2)}
-3Z_{1320}^{(2)}-3Z_{3102}^{(2)}\right),
\end{equation}
where $Z_{1320}^{(2)}=Z_{3102}^{(2)}$ due to the real-valued nature of the moments.
Since we also have
\begin{align}
Z_{2211}^{(2)}&=\frac{\gamma}{2}\left(4Z_{1122}^{(1)}
-Z_{0231}^{(1)}-Z_{2013}^{(1)}\right)=3\gamma^2,\nonumber\\
Z_{1320}^{(2)}&=\frac{\gamma}{2}\left(3Z_{0231}^{(1)}
-3Z_{1122}^{(1)}\right)=-3\gamma^2,
\end{align}
we obtain $Z_{3300}^{(3)}=15\gamma^3$ and
\begin{align}
\langle R^3\rangle&\approx 15\gamma^3 C_R^3=5!!\left(\frac{gkv_FT}{1+m_0^2}\right)^3.
\end{align}
Similarly, for the fourth-order coefficient of $Z_{4400}$, we have
\begin{equation}
Z_{4400}^{(4)}=\frac{\gamma}{4}\left(16Z_{3311}^{(3)}
-6Z_{2420}^{(3)}-6Z_{4202}^{(3)}\right),
\end{equation}
where $Z_{2420}^{(3)}=Z_{4202}^{(3)}$.
By using Eq.~(\ref{eq:recursion}) repeatedly, we obtain
\begin{align}
Z_{3311}^{(3)}&=\frac{\gamma}{3}\left(9Z_{2222}^{(2)}
-3Z_{1331}^{(2)}-3Z_{3113}^{(2)}\right)\nonumber\\
&=\frac{\gamma^2}{6}\left[9\left(4Z_{1133}^{(1)}-2Z_{0242}^{(1)}\right)
-6\left(3Z_{0242}^{(1)}-3Z_{1133}^{(1)}\right)\right]=15\gamma^3,\nonumber\\
Z_{2420}^{(3)}&=\frac{\gamma}{3}\left(8Z_{1331}^{(2)}-Z_{0440}^{(2)}
-6Z_{2222}^{(2)}\right)\nonumber\\
&=\frac{\gamma^2}{6}\left[8\left(3Z_{0242}^{(1)}-3Z_{1133}^{(1)}\right)+6Z_{0242}^{(1)}
-6\left(4Z_{1133}^{(1)}-2Z_{0242}^{(1)}\right)\right]=-15\gamma^3.
\end{align}
Then, we obtain $Z_{4400}^{(4)}=105\gamma^4$ and
\begin{align}
\langle R^4\rangle&\approx 105\gamma^4 C_R^4=7!!\left(\frac{gkv_FT}{1+m_0^2}\right)^4.
\end{align}
We can repeat similar derivations for higher values of $n$ and verify that in the short-time regime, it is generally valid that $Z_{nn00}^{(n)}=(2n-1)!!\gamma^n$ and
\begin{align}
\langle R^n\rangle&\approx (2n-1)!!\left(\frac{gkv_FT}{1+m_0^2}\right)^n.    
\end{align}

\section{Derivation of the analytical expression of the reflectance moments in the long-time regime, Eq.~(19)}
\label{app:asyml}

In the long-time regime, we anticipate that the time derivative of $Z_{abcd}$ will approach zero, under the assumption that the parameters $m_0$ and $g$ remain independent of time. Consequently, we can set the left-hand side of Eq.~(8) to zero, resulting in
a set of algebraic equations. From the structure of Eq.~(8), it becomes evident that the coefficient of $Z_{abcd}$ for $a\ne b$, when $a+c=b+d$, becomes complex. Therefore, these moments 
will continue to oscillate over time, unless they tend towards zero. Given that steady-state behavior is expected in the long-time limit, it is reasonable to assume that $Z_{abcd}$ for $a\ne b$ vanishes
under that condition. Based on these considerations, when $a=b=n$ and $c=d=0$, a straightforward equation can be deduced from Eq.~(8):
\begin{equation}
    Z_{nn00}=n\frac{1-B}{1+B}Z_{n-1,n-1,1,1}.
    \label{eq:ltl}
\end{equation}
We have also explicitly confirmed that $Z_{abcd}$ for $a\ne b$, when $a+c=b+d$, becomes zero through analytical solutions of the 
$n^2$ equations for any selected $n$. In the cases where $m_1=m_2=m_0$ as considered here, we have
$C_S=1$, $C_R=(1+B)/(1-B)$, 
and $S=\vert s\vert^2=1-R=1-C_R\vert r\vert^2$.
It then becomes straightforward to derive 
\begin{equation}
   Z_{n-1,n-1,1,1}= Z_{n-1,n-1,0,0}-C_RZ_{nn00}.
\end{equation}
By substituting this into Eq.~(\ref{eq:ltl}), we arrive at the recursion relation for the reflectance moments:
\begin{equation}
    \langle R^n\rangle=\frac{n}{n+1}\langle R^{n-1}\rangle,
\end{equation}
which can be easily solved to yield
\begin{equation}
    \langle R^n\rangle=\frac{1}{n+1}
\end{equation}
in the long-time regime.

\begin{figure}[H]
  \centering
  \includegraphics[width=9cm]{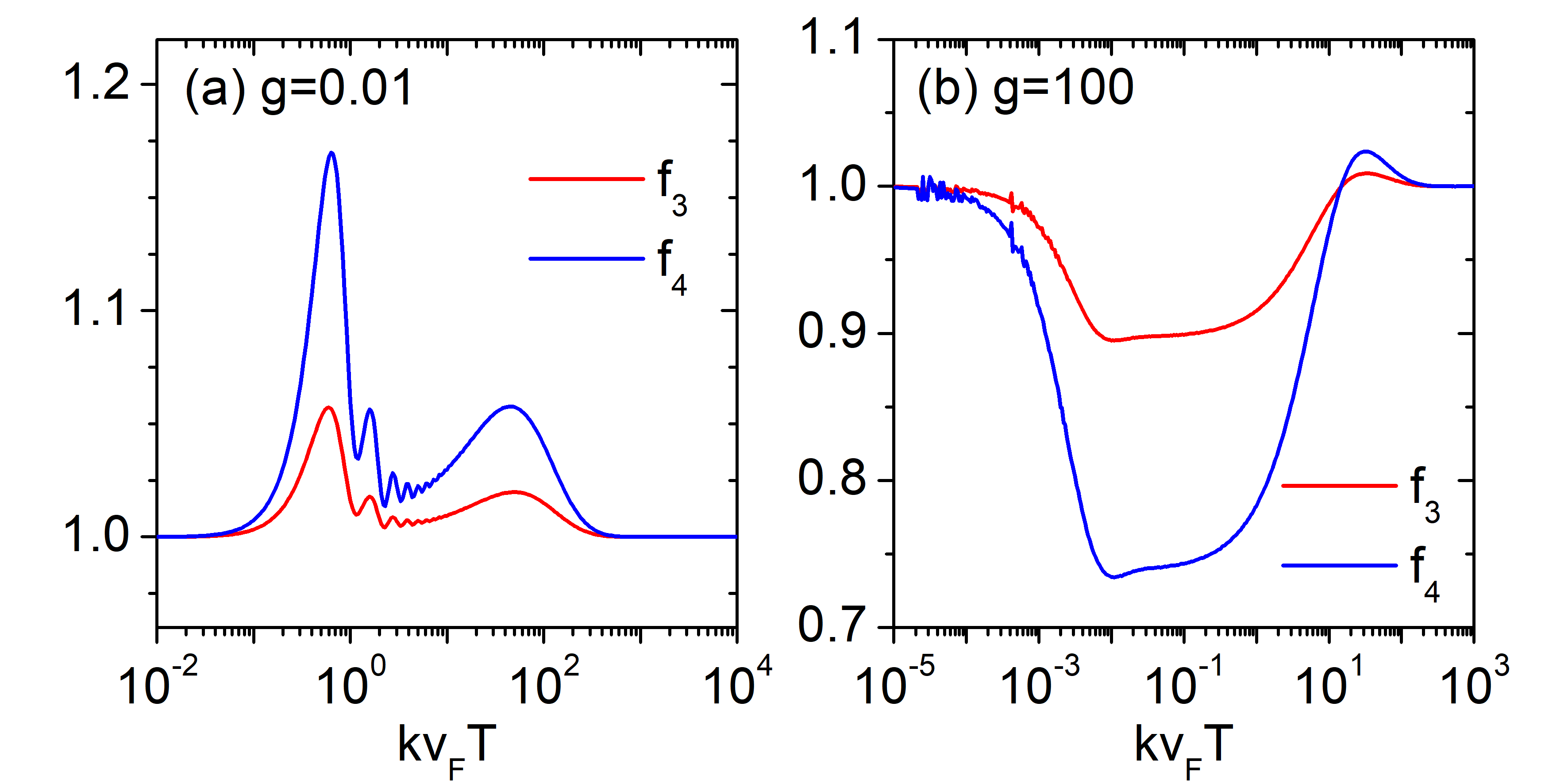}
  \caption{$f_3$ and $f_4$ for Model I plotted versus the normalized time interval $kv_F T$ on a linear-log scale, with $m_0=1$ and $g=0.01$ and 100.}
  \label{fig_bd}
\end{figure}

\section{Properties of the beta distribution and the demonstration of its satisfaction only in the short- and long-time regimes}
\label{app:demon}

The probability density function for the beta distribution is given by
\begin{equation}
P(x)=\frac{1}{B(\alpha,\beta)}x^{\alpha-1}(1-x)^{\beta-1},
\end{equation}
where $\alpha$ and $\beta$ are the shape parameters, and $B(\alpha,\beta)$ is the beta function. The raw moments $\langle x^n\rangle$ satisfy the recursion relation
\begin{equation}
\left\langle x^n\right\rangle=\frac{\alpha+n-1}{\alpha+\beta+n-1}\left\langle x^{n-1}\right\rangle,
\end{equation}
and the mean value of $x$ is given by $\langle x\rangle=\alpha/(\alpha+\beta)$. 

To ensure that the beta distribution is satisfied only in the short- and long-time regimes, we introduce the parameters $f_3$ and $f_4$, defined as follows:
\begin{equation}
f_3=\frac{\left\langle R^3\right\rangle}{\left\langle R^3\right\rangle_{b}},~
f_4=\frac{\left\langle R^4\right\rangle}{\left\langle R^4\right\rangle_{b}},
\end{equation}
where $\langle R^3\rangle_{b}$ and $\langle R^4\rangle_{b}$ are defined by
\begin{align}
&\left\langle R^3\right\rangle_{b}=\frac{\alpha+2}{\alpha+\beta+2}\left\langle R^{2}\right\rangle,~\left\langle R^4\right\rangle_{b}=\frac{\alpha+3}{\alpha+\beta+3}\left\langle R^{3}\right\rangle_{b}.
\end{align}
The effective shape parameters $\alpha$ and $\beta$ are obtained from the moments $\langle R\rangle$ and $\langle R^2\rangle$ using
\begin{align}
&\alpha=\left(\frac{\langle R\rangle-\langle R^2\rangle}{\left\langle R^2\right\rangle-\left\langle R\right\rangle^2}\right)\left\langle R\right\rangle,~\beta=\left(\frac{\langle R\rangle-\langle R^2\rangle}{\left\langle R^2\right\rangle-\left\langle R\right\rangle^2}\right)(1-\left\langle R\right\rangle).
\end{align}
When the beta distribution is satisfied, $f_3$ and $f_4$ should be identically equal to 1.
In Fig.~\ref{fig_bd}, we plot $f_3$ and $f_4$ versus the normalized time interval for the cases of weak $(g=0.01)$ and strong disorder ($g=100$) when $m_0=1$. In both cases, $f_3$ and $f_4$ converge to 1
in the short- and long-time regimes, while they deviate substantially from 1 in the intermediate time regime. 
This observation indicates that the beta distribution is satisfied only in the short- and long-time regimes.

\begin{figure}
  \centering
  \includegraphics[width=9cm]{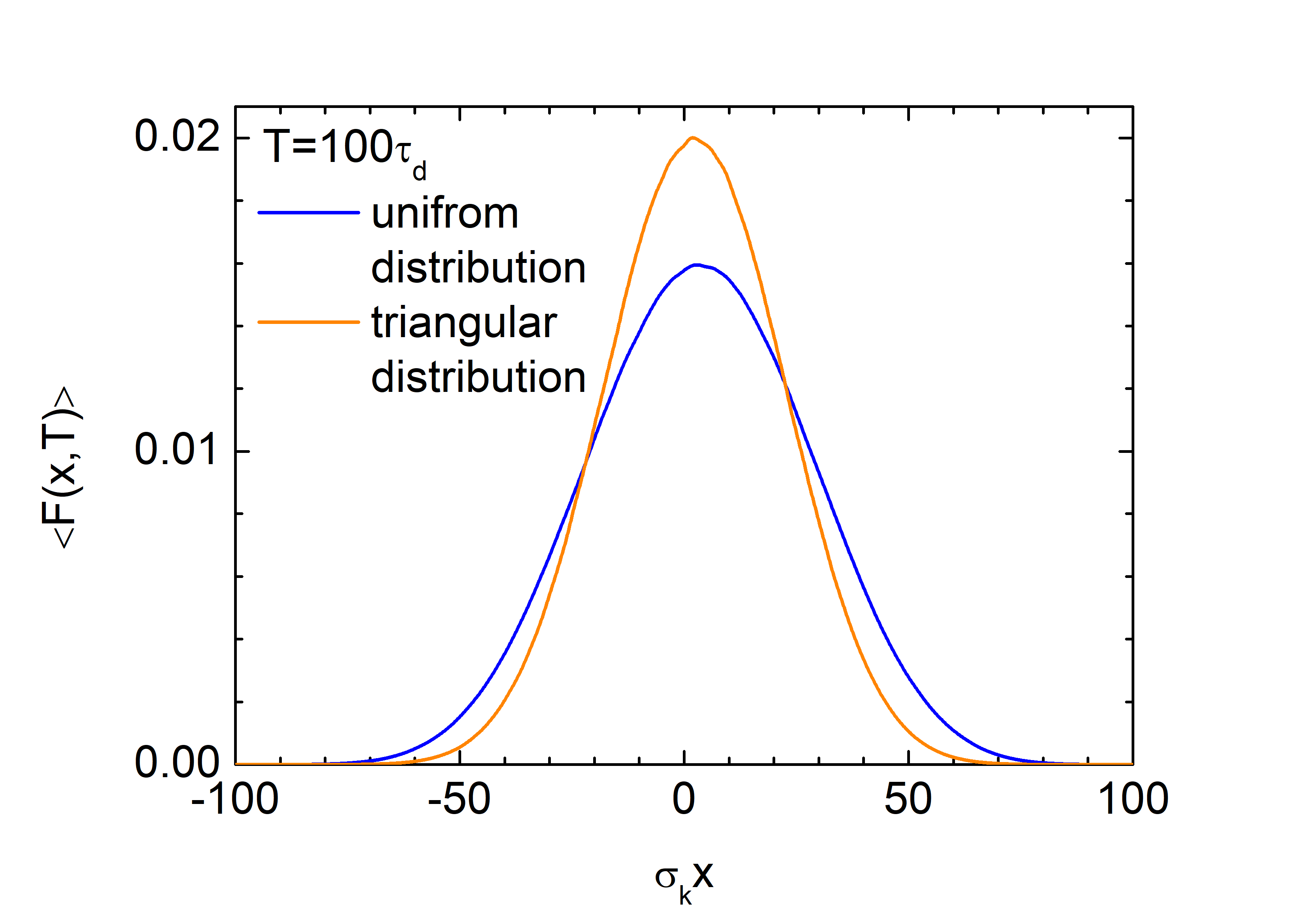}
  \caption{Spatial distribution of the disorder-averaged field intensity of a pulse $\left\langle F(x, T)\right\rangle$ at $T=100\tau_d$, obtained for a triangular distribution of the random mass, is compared with the corresponding quantity for a uniform distribution, as shown in Fig.~6(g). Disorder averaging is performed over $10^5$ independent realizations.}
  \label{gg1}
\end{figure}

\begin{figure}
  \centering
  \includegraphics[width=9cm]{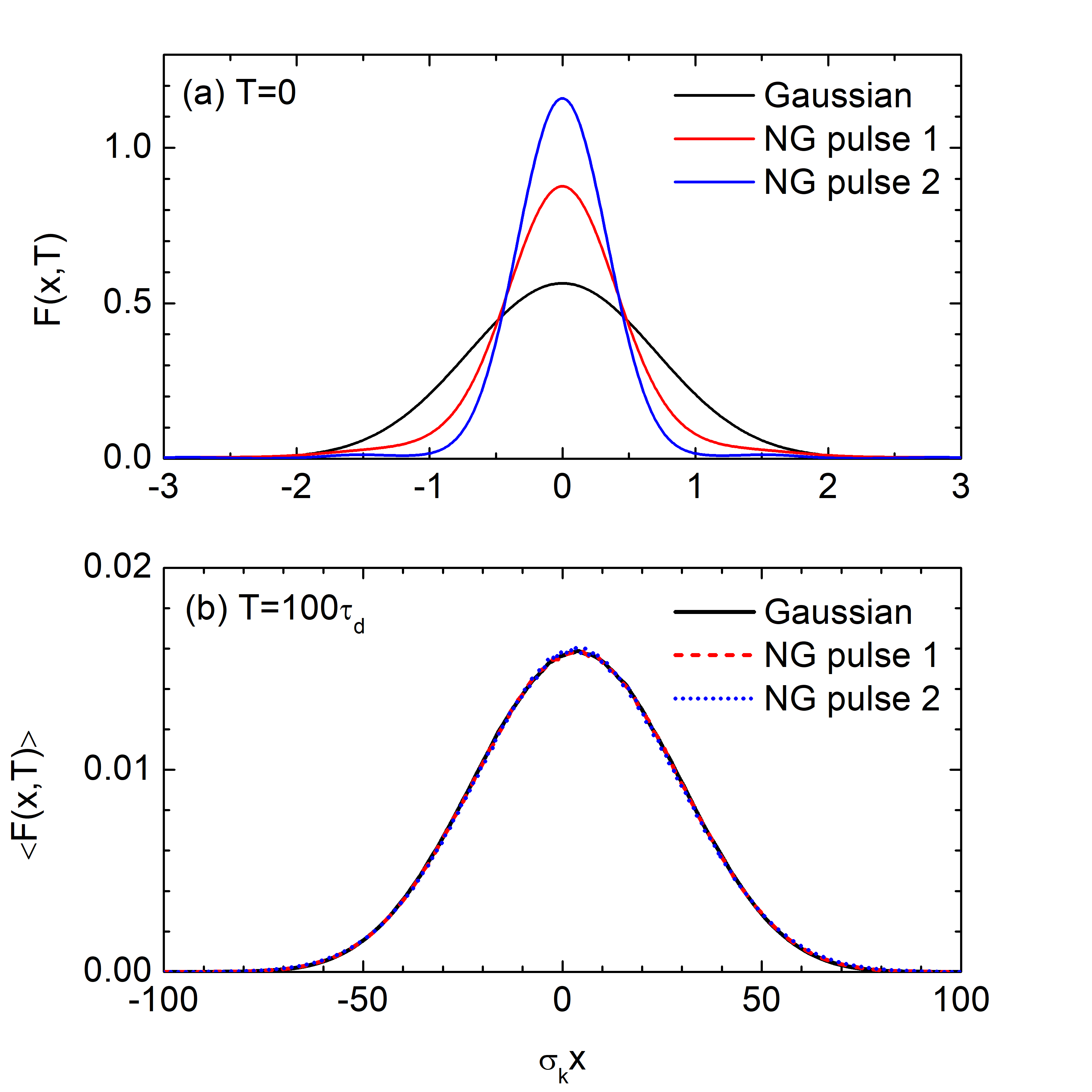}
  \caption{(a) A Gaussian pulse and two non-Gaussian (NG) pulses used as initial shapes in the time-evolution calculation. NG pulses 1 and 2
  are defined by $D_1$ and $D_2$ in Eq.~(\ref{eq:ngp}), respectively. (b) Spatial distributions of $\left\langle F(x, T)\right\rangle$ at $T=100\tau_d$, obtained by averaging over $10^5$ realizations for a uniform distribution of the random mass, corresponding to the initial pulse shapes shown in (a).}
  \label{gg2}
\end{figure}

\section{Universal Gaussian pulse shape in the long-time regime}

In this section, we demonstrate that pulse shapes at long times are well-described by Gaussian functions, regardless of the probability distribution of the random mass or the initial pulse shape. In Fig.~\ref{gg1}, we compare the spatial distribution of the disorder-averaged field intensity of a pulse, 
$\left\langle F(x, T)\right\rangle$, at $T=100\tau_d$, obtained for a triangular distribution of the random mass, with the corresponding quantity for a uniform distribution, as shown in Fig.~6(g). The probability density function of the mass in the triangular distribution is assumed to be defined in the range $[-2, 2]$, with the maximum value of 0.5 at the origin. Disorder averaging is performed over $10^5$ independent realizations. We find that the pulse shapes are precisely Gaussian in both cases, although their widths at $T=100\tau_d$ differ.

In Fig.~\ref{gg2}, we investigate the dependence of the pulse shape in the long-time regime on the initial pulse shape. Three pulses with different initial shapes are considered, including the Gaussian pulse shown in Fig.~6(b) and two non-Gaussian pulses defined as in Eq.~(20), each with their respective momentum distribution functions, $D_{1}(k_x)$ and $D_{2}(k_x)$, given by
\begin{align}
&D_1(k_x)=e^{-\frac{\left(k_x-k_c\right)^2}{2\sigma_k^2}}
+e^{-\frac{\left(k_x-k_c\right)^2}{10\sigma_k^2}}
+e^{-\frac{5\left(k_x-k_c\right)^2}{\sigma_k^2}},\nonumber\\
&D_2(k_x)=e^{-\frac{\left(k_x-k_c\right)^2}{5\sigma_k^2}}
+e^{-\frac{\left(k_x-k_c\right)^2}{20\sigma_k^2}}
+e^{-\frac{5\left(k_x-k_c\right)^2}{\sigma_k^2}}.
\label{eq:ngp}
\end{align}
We find that in the long-time regime, the pulse shapes converge to the same Gaussian function
in all three cases.

\end{document}